\documentclass[aps,twocolumn,preprintnumbers,amsmath,amssymb,nofootinbib,superscriptaddress,notitlepage]{revtex4-1}

\usepackage{epsfig}
\usepackage[utf8]{inputenc}
\usepackage{color}

\begin{document}

\title{Subleading contributions to the decay width of the $T_{cc}^+$ tetraquark}

\author{Mao-Jun Yan}
\affiliation{School of Physics, Beihang University, Beijing 100191, China} 
\affiliation{CAS Key Laboratory of Theoretical Physics, 
  Institute of Theoretical Physics, \\
  Chinese Academy of Sciences, Beijing 100190, China}

\author{Manuel Pavon Valderrama}\email{mpavon@buaa.edu.cn}
\affiliation{School of Physics, Beihang University, Beijing 100191, China} 

\date{\today}


\begin{abstract} 
  \rule{0ex}{3ex}
  Recently the LHCb collaboration has announced the discovery of
  the $T_{cc}^+$ tetraquark.
  Being merely a few hundred ${\rm keV}$ below the $D^{*+} D^0$ threshold,
  the $T_{cc}^+$ is expected to have a molecular component,
  for which there is a good separation of scales
  that can be exploited to make reasonably accurate theoretical
  predictions about this tetraquark.
  Independently of its nature, the most important decay channels will be
  $D^+ D^0 \pi^0$, $D^0 D^0 \pi^+$ and $D^+ D^0 \gamma$.  
  Its closeness to threshold suggests that the mass and particularly
  the width of the $T_{cc}^+$ tetraquark depend on the resonance profile.
  While the standard Breit-Wigner parametrization generates a $T_{cc}^+$
  that is too broad for current theoretical calculations to reproduce,
  a three-body unitarized Breit-Wigner shape reveals instead
  a decay width ($\Gamma_{\rm pole} = 48\pm 2\,{}^{+0}_{-12}\,{\rm keV}$)
  consistent with theoretical expectations.
  Here we consider subleading order contributions to the decay amplitude,
  which though having at most a moderate impact in the width
  still indicate potentially significant differences with
  the experimental width that can be exploited to
  disentangle the nature of the $T_{cc}^+$.
  Concrete calculations yield $\Gamma^{\rm LO} = 49 \pm 16\,{\rm keV}$
  and $\Gamma^{\rm NLO} = 58^{+7}_{-6}\,{\rm keV}$, though we expect
  further corrections to the ${\rm NLO}$ decay widths from asymptotic
  normalization effects.
  We find that a detailed comparison of the ${\rm NLO}$ total and
  partial decay widths with experiment suggests the existence of
  a small (but distinguishable from zero) non-molecular
  component of the $T_{cc}^+$. 
\end{abstract}

\maketitle

\section{Introduction}

%
The LHCb collaboration has recently observed~\cite{LHCb:2021vvq}
a tetraquark in the $D^0 D^0 \pi^+$ mass spectrum.
The Breit-Wigner parameters of this tetraquark, the $T_{cc}^+$, are
\begin{eqnarray}
  \delta m_{\rm BW} &=& - 273 \pm 61 \pm 5 \, {}^{+11}_{-14} \,{\rm keV} \, ,
  \label{eq:m-BW} \\
  \Gamma_{\rm BW} &=& 410 \pm 165 \pm 43 \, {}^{+18}_{-38} \,{\rm keV} \, , 
\end{eqnarray}
where the mass difference is with respect to the $D^{*+} D^0$ threshold.
Alternatively, if the data are analyzed with a resonance
profile more suitable to the closeness of the $T_{cc}^+$ to the
$D^{*+} D^0$ threshold~\cite{LHCb:2021auc}, the parameters of
the $T_{cc}^+$ pole turn out to be
\begin{eqnarray}
  \delta m_{\rm pole} &=& - 360 \pm 40 {}^{+4}_{-0} \,{\rm keV} \, ,
  \label{eq:m-pole} \\
  \Gamma_{\rm pole} &=& 48 \pm 2 {}^{+0}_{-12} \,{\rm keV} \, . 
\end{eqnarray}
Of course the question is what is the nature of this state, where
the two contending explanations are a compact tetraquark or
a loosely bound $D^{*+} D^0$-$D^{*0} D^+$ system.

Actually, there is a long list of predictions of a $cc \bar{u} \bar{d}$
state with $J=1^+$ and $I=0$, beginning with the pioneering realization
by Zouzou et al.~\cite{Zouzou:1986qh} that this tetraquark could be
below the $D^* D$ threshold, followed by a large series of
works till nowadays~\cite{Carlson:1987hh,Silvestre-Brac:1993zem,Semay:1994ht,Pepin:1996id,Janc:2004qn,Navarra:2007yw,Yang:2009zzp,Karliner:2017qjm,Wang:2018poa}.
The predictions of $J= 1^+$, $I=0$ $D^* D$ and $D^* D^*$ bound states
(for which heavy quark-spin symmetry predicts identical
potentials~\footnote{The situation
  is completely analogous to the $Z_b(10610)$ and
  $Z_b(10650)$~\cite{Bondar:2011ev}, or to the $X(3872)$ and
  its hypothetical $J^{PC} = 2^{++}$ $X(4012)$
  partner~\cite{Nieves:2012tt}. However, for the $D^* D^*$ a similar
  caveat apply as for the $X(4012)$~\cite{Cincioglu:2016fkm}:
  the actual location of the compact $cc \bar{u} \bar{d}$ / $c\bar{c}$ states
  might make the higher mass partner disappear.})
are in contrast somewhat more
recent, with Manohar and Wise~\cite{Manohar:1992nd} and
T\"ornqvist~\cite{Tornqvist:1993ng} considering it unlikely (from pions alone),
but then Ericson and Karl~\cite{Ericson:1993wy} realizing that this conclusion
might change if other meson exchanges are considered, an observation later
confirmed in~\cite{Molina:2010tx} for $D^* D^*$, in \cite{Li:2012ss}
for $D^* D$ (corresponding to the $T_{cc}^+$),
in~\cite{Liu:2019stu} for $D^* D$ and $D^* D^*$,
in~\cite{Ding:2020dio,Xu:2017tsr} for $D^*D$, etc.
(plus the attention this hypothesis
has received~\cite{Li:2021zbw,Wu:2021kbu,Agaev:2021vur,Chen:2021vhg,Dong:2021bvy}
after the observation of the $T_{cc}^+$).
Here it is worth noticing that there might be up to three states
with the quantum numbers of the $T_{cc}^+$ (with the molecular
ones usually close to threshold).

In view of the aforementioned theoretical landscape, there are reasons
to believe the two hypotheses: the molecular and tetraquark explanations
are not mutually exclusive and the $T_{cc}^+$
could be a superposition of both.
The molecular component of the $T_{cc}^+$ has the theoretical advantage of
being a shallow bound state, presumably with a good separation
of scales between its long-range and short-range components.
This in turn allows us to use the existent theoretical toolbox
for shallow bound states~\cite{vanKolck:1998bw,Chen:1999tn},
from which in principle it would be possible to make predictions
accurate enough as to analyze its structure.

In this regard, the decay width of the $T_{cc}^+$ is particularly important
(and indeed it has already received due
attention~\cite{Meng:2021jnw,Ling:2021bir,Feijoo:2021ppq}):
if the experimental measurements and theoretical predictions
are on par with each
other in terms of accuracy, we will be able to determine whether the
$T_{cc}^+$ is purely molecular or compact, or what is the degree of
admixture between these two explanations.
Given small enough uncertainties, a calculated decay width
that is too small or too large in comparison with the experiment might
point out to (or even determine) the existence of physics beyond
the naive molecular explanation, like a tetraquark component
or unobserved states.
However, this might prove difficult: the wave function of a tetraquark
close to the $D^* D$ threshold might be indistinguishable
from that of two separate $D^*$ and $D$ mesons, as noted
in~\cite{Janc:2004qn}, which already considered
the possibility of a tetraquark lying between
the $D^* D$ and $D D\pi$ thresholds (see also
the discussion in~\cite{Qin:2020zlg}).

\section{Decay channels:}

%
The $T_{cc}^+$ decay width is expected to be saturated by its strong and
electromagnetic decays, which are in principle limited to three
possibilities: $T_{cc}^+ \to D^+ D^0 \pi^0$, $T_{cc}^+ \to D^0 D^0 \pi^+$
and $T_{cc}^+ \to D^+ D^0 \gamma$.

However this is not necessarily the whole story: if the predicted
$cc \bar{u} \bar{d}$ tetraquark happens to be a different
state than the $T_{cc}^+$ but with a lower mass, which we might call $T_{cc}^{'+}$
for concreteness, we will have to add up to two new decay channels:
$T_{cc}^+ \to T_{cc}^{'+} \gamma$ (M1 magnetic and E2 quadrupole
transitions) and $T_{cc}^+ \to T_{cc}^{'+} \pi^0$, of which
the second requires isospin breaking (e.g. stemming from
the isospin breaking in the mass of the $D^{*+} D^0$ and
$D^+ D^{*0}$ channels) and a compact $T_{cc}^{'+}$
located close or below the $D^+ D^0$ threshold.
This last condition is more difficult to meet as there are less
predictions of a $cc \bar{u} \bar{d}$ state close or below the $DD$
threshold~\cite{Feng:2013kea,Deng:2018kly,Yang:2019itm,Gao:2020ogo}
(as to allow some phase space for $T_{cc}^+ \to T_{cc}^{'+} \pi^0$)
than between the $D^* D$ and $D D$ thresholds~\cite{Silvestre-Brac:1993zem,Gelman:2002wf,Vijande:2003ki,Lee:2009rt,Ikeda:2013vwa,Junnarkar:2018twb}.
A different variation over this idea --- the possibility of a $DD$ bound sate,
$\tilde{T}_{cc}^+$, and its potential effect on the $T_{cc}^+$ decay width ---
has been recently explored in Ref.~\cite{Fleming:2021wmk}.

The most straightforward calculation of the $T_{cc}^+$ decay width into pions
involves sandwiching the $D^* \to D \pi$ one-body decay operators between
the initial and final wave
functions~\cite{Meng:2021jnw,Ling:2021bir,Feijoo:2021ppq},
in which case the total decay width of a molecular $T_{cc}^+$ falls short of
the Breit-Wigner width~\cite{LHCb:2021vvq}, but agrees well with the width
from the improved resonance profile introduced in~\cite{LHCb:2021auc}
(which is in turn consistent with the well known fact that the Breit-Wigner
parametrization will lead to distortions for two-body states close to threshold~\cite{Hanhart:2015cua,Dong:2020hxe}).
Here we include a series of subleading order effects, including two-body decay
operators and rescattering effects in the final $DD$ pair, which 
refine the aforementioned theoretical estimations and might
allow to eventually disentangle the molecular and
non-molecular components of the $T_{cc}^+$.

\section{Power counting:}

%
Effective field theories (EFTs) are expansions in terms of the ratio 
$Q/M$, with $Q$ and $M$ characteristic soft and hard scales of
the system at hand.
If the $T_{cc}^+$ is molecular, its natural momentum scale $Q$ is given by
the wave number of its $D^{*+} D^0$-$D^{*0} D^+$ components, i.e.
$23-26$ and $57-59\,{\rm MeV}$, respectively, depending on
whether we use $\delta m_{\rm BW}$ or $\delta m_{\rm pole}$,
see Eqs.~(\ref{eq:m-BW}) and (\ref{eq:m-pole}).
The ratio of these two scales with respect to the pion mass is about
$0.18$ and $0.42$, from which it would be perfectly possible to consider
the pion mass as a heavy scale $M \sim m_{\pi}$ in a first approximation.
If we consider the strong decay products of the $T_{cc}^+$, the maximum momentum
and energy of the final pion are about $40$ and $6\,{\rm MeV}$,
which are again small in comparison with the pion mass.
The situation is less clear with the momentum of the final $DD$ pair, which
can reach $100\,{\rm MeV}$: however, one pion exchange does not happen
in this system, with the longest range piece of the $DD$ potential
being the two-pion exchange football diagram, with a range of
$2 m_{\pi}$. From this, the ratio of scales for the final $DD$
system is $0.37$.
At this point it is worth noticing that a pion exchanged between a $D^* D$ and
a $D D^*$ initial and final state is almost on mass shell and will {
  in principle follow naive dimensional analysis (NDA)
  as its power counting (also referred to as
  Weinberg's counting~\cite{Weinberg:1990rz,Weinberg:1991um},
  which was originally formulated for the two-nucleon system
  but can be applied to other non-relativistic
  two-hadron systems as well).}
{This conclusion changes though} once
we consider the relatively large momentum scale at which pions become
non-perturbative {in the two-charmed meson
  system}~\cite{Valderrama:2012jv}. 
In summary, an effective field theory description in which the pion mass is
considered a hard scale is expected to have a convergence parameter
in the range $Q/M \sim 0.2-0.4$.

\begin{figure}
  \includegraphics[width=45mm]{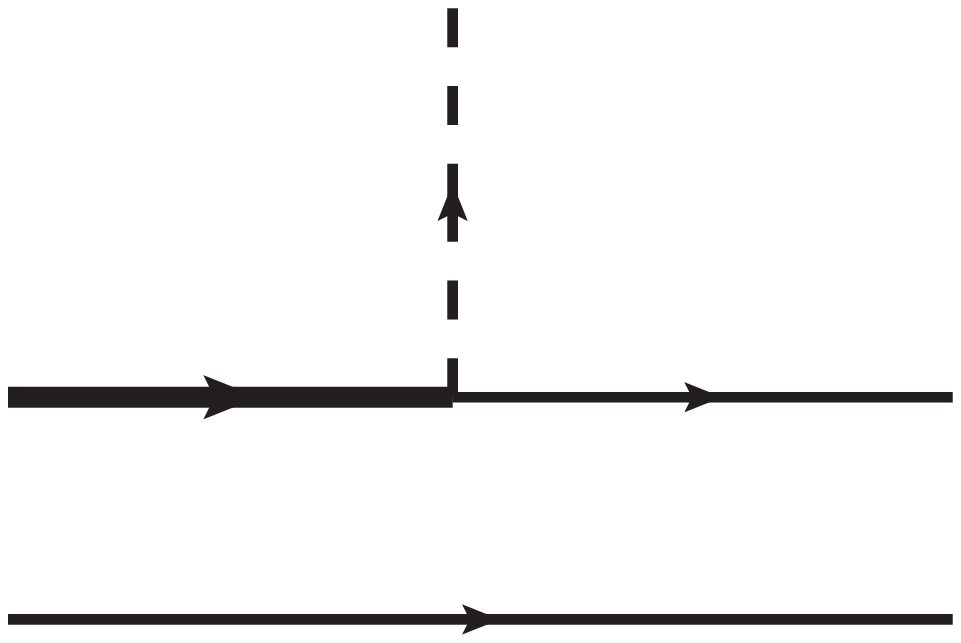} 
  \includegraphics[width=45mm]{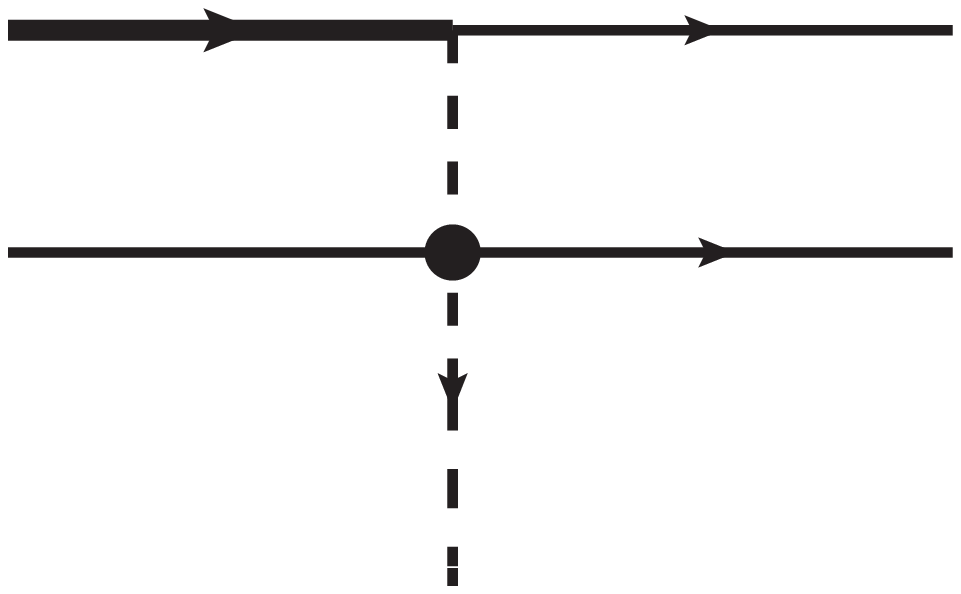} 
  \includegraphics[height=35mm]{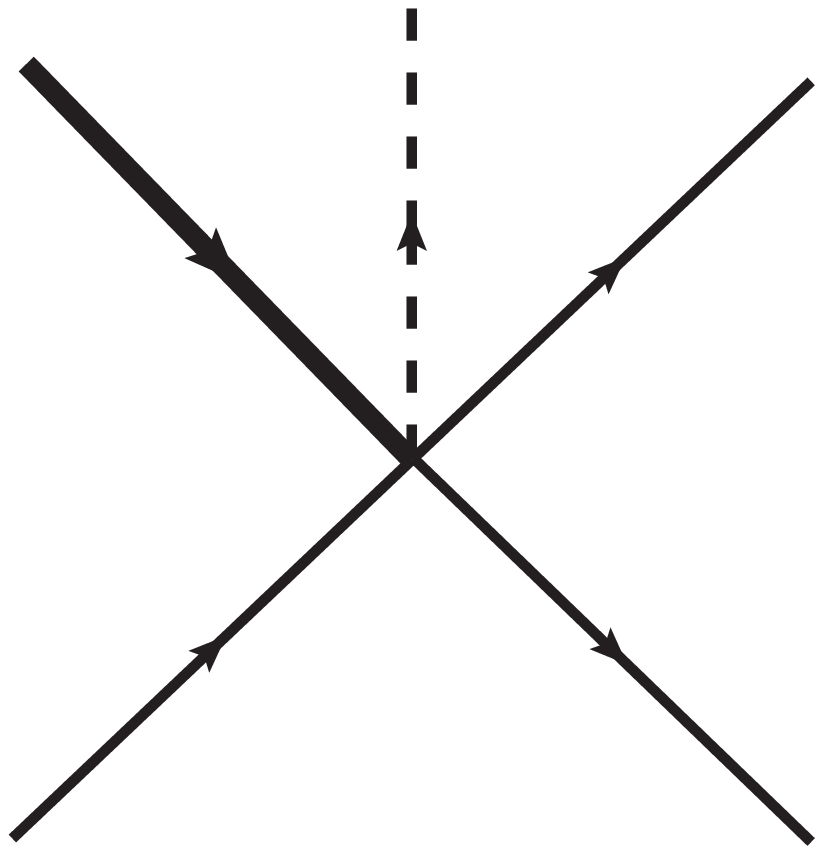} 
\caption{
  Lowest-order operators involved in the $T_{cc} \to DD\pi$ decay:
  the one-body decay operator is of order $Q^{-2}$, while
  the seagull and contact two-body currents are naively 
  of order $Q^0$ and $Q^{1}$ but get promoted to $Q^{-1}$
  and $Q^0$, respectively.
}
\label{fig:operators}
\end{figure}

With this, if we consider the strong decays of the $T_{cc}^+$ and
the diagrams in Fig.~\ref{fig:operators}, their counting will be
\begin{itemize}
\item[(a)] the one-body decay diagram is order $Q^{-2}$ and, being the lowest
  order one, it is leading order (${\rm LO}$),
\item[(b)] {in the Weinberg counting} the seagull diagram is $Q^0$,
  but in the decay of the $T_{cc}^+$ the Weinberg-Tomozawa term is proportional
  to $m_{\pi}$ which we count as a hard scale. Thus this diagram is
  promoted to $Q^{-1}$ and is next-to-leading order (${\rm NLO}$),
\item[(c)] the contact-range two-body operator is naively $Q^1$, but if the
  $T_{cc}^+$ is a bound state and applying the arguments
  of Ref.~\cite{PavonValderrama:2014zeq}
  for the counting of two-body operators, it will be promoted to $Q^0$
  and will be next-to-next-to-leading order (${\rm N^2LO}$).
  
  Indeed, following the logic in Ref.~\cite{PavonValderrama:2014zeq},
  if we apply renormalization group invariance to the contact-range
  two-body operator, we obtain
  \begin{eqnarray}
     \frac{d}{d R_c}\,\langle \Psi(DD) | \hat{\mathcal{O}}^{2B}_C | \Psi(T_{cc}^+) \rangle &\propto& \nonumber \\
     \frac{d}{d R_c}\,\left[ \vec{\epsilon}_1 \cdot \vec{q}\,
       \frac{C_{2B}(R_c)}{R_c} \right] &=& 0\, ,
  \end{eqnarray}
  where $|\Psi(T_{cc}^+)\rangle$ and $|\Psi(DD)\rangle$ are the initial and
  final two-meson wave functions, $R_c$ is a cutoff radius,
  $C_{2B}$ the contact-range coupling, $\vec{\epsilon}_1$
  the polarization of the $T_{cc}^+$ tetraquark and
  $\vec{q}$ the momentum of the pion.
  The $R_c^{-1}$ factor in front of $C_{2B}(R_c)$ comes from the tetraquark wave
  function, which scales as $\langle r | \Psi(T_{cc}^+) \rangle \propto 1/r$
  at short distances.
  This factor also implies that in the infrared limit ($R_c^{-1} \sim Q$),
  the $C_{2B}$ coupling is proportional to $1/Q$ and thus enhanced by one
  order with respect to NDA.
\end{itemize}
It is worth noticing that the appearance of the contact-range two-body operator
sets the limit of predictability of the EFT: at ${\rm N^2LO}$ this contact
can be calibrated to reproduce the $T_{cc}^+$ decay width into pions,
which means that the decay width becomes the input of the theory
(instead of its output, which is what we want).
For comparison purposes, this counting has a few similarities
with X-EFT~\cite{Fleming:2007rp} and a very significant
difference in what regards the counting of
pion exchanges (which we count at least as ${\rm N^2LO}$,
as will be explained later, see the discussion below Eq.~(\ref{eq:V-SL})).
We find it also interesting to comment on Ref.~\cite{Fleming:2021wmk},
which proposes an EFT description of the $T_{cc}^+$ decays
when there is a bound state in the final $DD$ state:
if this were to be the case, the final $DD$ wave function will
behave in exactly the same way as the initial $T_{cc}^+$ one,
i.e. $\langle r |\Psi(DD)\rangle \propto 1/r$,
which will result in a $1/Q^2$ enhancement of
the contact-range two-body operator,
which will then enter at ${\rm NLO}$.
In this case, EFT will only be able to predict the $T_{cc}^+$ decay
at ${\rm LO}$ (instead of ${\rm NLO}$ as we propose here).

Even if the ${\rm NLO}$ limitation and the potentially slow convergence
parameter look disappointing, they are indeed more than enough for
the current situation: the relative uncertainty in the experimental
decay width is $0.43$ for the standard Breit-Wigner parametrization
and $0.25$ for the unitarized Breit-Wigner.
This naively indicates that either a ${\rm LO}$ or ${\rm NLO}$
calculation will be enough to match it, but at which order
this exactly happens is not completely obvious a priori:
EFT arguments allow for the existence of numerical factors of
$\mathcal{O}(1)$, which might subvert the original
power counting expectations.
If we ignore these numerical factors and consider 
the expansion parameter to lie between $0.18-0.42$, we find that the
uncertainty in the ${\rm LO}$ and ${\rm NLO}$ decay widths will be
\begin{eqnarray}
  \frac{\Delta \Gamma^{\rm LO}}{\Gamma^{\rm LO}} \sim 0.18-0.42 
  \quad \mbox{and} \quad
  \frac{\Delta \Gamma^{\rm NLO}}{\Gamma^{\rm NLO}} \sim 0.03-0.18 \, ,
  \nonumber \\
  \label{eq:naive-EFT-convergence}
\end{eqnarray}
which indicates that a ${\rm NLO}$ calculation is necessary to be fully
competitive with experiment, particularly if we want to match
the accuracy of~\cite{LHCb:2021auc}.
As we will see, calculations of the decay width will turn out to be compatible
with the average estimations of the EFT convergence.
Thus it happens that all the pieces fit together to put the current limit
at ${\rm NLO}$, as it is simply not possible to achieve a better accuracy 
at ${\rm N^2LO}$ where the decay width is no longer a prediction.

\section{Decay amplitudes}

For the decay of the $T_{cc}^+$ into $DD\pi$
we will consider a decay amplitude in the form
\begin{eqnarray}
  \langle D D (\vec{p}\,') \pi^c  | H | D^* D (\vec{p}) \rangle =
  A^c (\vec{p}\,', \vec{p}, \vec{q}) \, ,
\end{eqnarray}
where $c$ is the isospin index of the outgoing pion, $\vec{q}$ its momentum
and $\vec{p}$ ($\vec{p}\,'$) the center-of-mass relative momentum of
the incoming (outgoing) $D^* D$ ($D D$) system.
This amplitude will be sandwiched between the initial and final state
wave functions
\begin{eqnarray}
  \langle A^c \rangle = \langle D D (\vec{k}) | A^c | T_{cc}^+ \rangle \, ,
\end{eqnarray}
then inserted into Fermi's golden rule to obtain the decay width
\begin{eqnarray}
  \Gamma (T_{cc}^+ \to D D \pi) &=& 2 \pi \,
  \int \frac{d^3 \vec{k}}{(2 \pi)^3}\,
  \frac{d^3 \vec{q}}{(2 \omega) (2 \pi)^3} \nonumber \\ &\times&
  \delta (\omega + \frac{k^2}{2 \mu_{DD}} + \frac{q^2}{2 m_{DD}} - \Delta)\,
  \overline{\left| \langle A^c \rangle \right|^2} \, , \nonumber \\
  \label{eq:fermi-golden-rule-DDpi}
\end{eqnarray}
where $\vec{k}$ is the center-of-mass momentum of the $DD$ pair,
$\vec{q}$ refers to the momentum of the outgoing pion and
$\omega = \sqrt{m_{\pi}^2 + q^2}$ to its energy
(with $m_{\pi}$ the pion mass),
$\mu_{DD}$ and $m_{DD}$ are the reduced and total mass of the final $DD$ pair
and $\overline{\left| \langle A^c \rangle \right|^2}$ represents
the sum over final states and average over initial states.
The amplitude $A^c$ is represented by the diagrams
in Fig.~\ref{fig:operators}, the evaluation of which yields
\begin{eqnarray}
  A_c &=& \frac{g_1}{\sqrt{2} f_{\pi}} \Big[ \vec{\epsilon}_1 \cdot \vec{q}\,
    \,\tau_{1c}\,(2 \pi)^3 \delta^{(3)} (\vec{p}\,' - \vec{p} + \frac{\vec{q}}{2})
    \nonumber \\
    &+& i (\vec{\tau}_1 \times  \vec{\tau}_2)_c\,\frac{m_{\pi}}{f_{\pi}^2}\,
    \frac{\vec{\epsilon}_1 \cdot (\vec{p}\,' - \vec{p} - \frac{\vec{q}}{2})}
         {\mu_{\pi}^2 + (\vec{p}\,' - \vec{p} - \frac{\vec{q}}{2})^2}\, \Big] \, ,
\end{eqnarray}
where $\vec{p}$, $\vec{p}\,'$ are the relative momenta of the incoming and
outgoing $D^* D$ and $D D$ systems, $\vec{q}$ and $c$ the momentum and
the isospin index (in the Cartesian basis) for the outgoing pion,
$\tau_{i}^c$ is the isospin operator (a Pauli matrix) for the pion
as applied to vertex $i=1,2$, $f_{\pi} \simeq 130\,{\rm MeV}$
the pion weak decay constant, $g_1$ the axial coupling
for the charmed mesons, $m_{\pi}$ the pion mass and
$\mu_{\pi}^2 = m_{\pi}^2 - (m(D^*) - m(D))^2$ the effective pion mass
for the in-flight pion, which can be on-shell and which 
we simplify to $\mu_{\pi} = 0$ from now on~\footnote{Actually,
  $\mu_{\pi}^2 \leq 0$, which means that we can interpret
  the two-body operator as the rescattering of
  the outgoing pion with the second charmed
  meson (indeed, this is how this operator is interpreted
  in X-EFT~\cite{Dai:2019hrf}).
  By taking the $\mu_{\pi} = 0$ limit we are effectively
  considering that this rescattering happens at zero energy,
  which is a good approximation taking into account
  that the maximum momentum of the pion
  is about $35-40\,{\rm MeV}$.}.
Besides, in the Weinberg-Tomozawa vertex we have made the simplification
that the energy of the incoming and outgoing pion is $m_{\pi}$ (we notice
that changing it to $\omega_{\pi} = \sqrt{m_{\pi}^2 + {\vec{q}\,}^2}$
has a negligibly small effect).
For the initial center-of-mass momentum coordinates we have also ignored
the mass difference between the $D$ and $D^*$ mesons.

We now evaluate the decay operator between the initial and final states.
If we assume wave functions of the type
\begin{eqnarray}
  \langle \vec{x} | DD(\vec{k}) \rangle = e^{i \vec{k} \cdot \vec{x}}\,| D D \rangle
  \,\,\, \mbox{and} \,\,\,
  \langle \vec{x} | T_{cc} \rangle = \psi(\vec{x})\,| D^* D \rangle \, ,
  \nonumber \\
\end{eqnarray}
the matrix element of the decay amplitude can be explicitly evaluated as
follows
\begin{eqnarray}
  \langle A_c \rangle &=& \frac{g_1}{\sqrt{2} f_{\pi}}
  \Big[ \vec{\epsilon}_1 \cdot \vec{q}\,
    \langle \tau_{1c} \rangle \, I_{\rm 1B}(\vec{k},\vec{q})
    \nonumber \\
    &+& i (\vec{\tau}_1 \times  \vec{\tau}_2)_c\,
    \frac{m_{\pi}}{f_{\pi}^2}\,i \vec{\epsilon}_1 \cdot 
    \vec{I}_{\rm 2B}(\vec{k},\vec{q})
    \Big] \, ,
\end{eqnarray}
where the one- and two-body integrals take the form
\begin{eqnarray}
  I_{\rm 1B}(\vec{k},\vec{q}) &=&
  \int \, d^3 \vec{x} \, \psi(\vec{x}) e^{-i (\vec{k} - \frac{\vec{q}}{2}) \cdot
    \vec{x}} \, , \\
  \vec{I}_{\rm 2B}(\vec{k},\vec{q}) &=& 
  \int d^3 \vec{x} \psi(\vec{x})\,
  \vec{\nabla}_x
  \left[ \frac{1}{4 \pi {|\vec{x}|}} \right]\,
   e^{-i (\vec{k} + \frac{\vec{q}}{2}) \cdot \vec{x}} \, . \nonumber \\
\end{eqnarray}

These expressions can be further simplified by assuming the $T_{cc}$ to be
an S-wave bound state
\begin{eqnarray}
  \langle \vec{r}\, | T_{cc} \rangle &=& \frac{1}{\sqrt{4\pi}}\,\frac{u(r)}{r}\,
  | D^* D \rangle \, , 
\end{eqnarray}
and by expanding the decay amplitude in partial waves
\begin{eqnarray}
  \langle A(\vec{k},{\vec{q}}) \rangle &=&
  \vec{\epsilon}_1 \cdot \vec{q}\, A_{SP}(k,q) +
  \vec{\epsilon}_1 \cdot \vec{k}\, A_{PS}(k,q) 
  \nonumber \\
  &+& (\mbox{D-waves and higher}) \, ,
\end{eqnarray}
where we will ignore contributions in which the final $DD$ pair has orbital
angular momentum $L \geq 2$.
After a few manipulations we arrive at
\begin{eqnarray}
  \langle A_c \rangle 
  &=&
  \frac{g_1}{\sqrt{2} f_{\pi}}\,\Big[
    \vec{\epsilon}_1 \cdot \vec{q}\,
    \langle \tau_{1c} \rangle \, I_{00}(k,q)
    \nonumber \\
    &-& \frac{m_{\pi}}{f_{\pi}^2} \,
    \langle i (\vec{\tau}_1 \times  \vec{\tau}_2)_c \rangle
    \Big( \vec{\epsilon}_1 \cdot \frac{\vec{q}}{2}\,I_{01}(k,q)
    \nonumber \\
    && \qquad 
    + \vec{\epsilon}_1 \cdot \vec{k}\,I_{10}(k,q) \Big)
    \Big] \, ,
  \label{eq:Ac-matrix-element}
\end{eqnarray}
where the integrals $I_{00}$, $I_{01}$ and $I_{11}$ are given by
\begin{eqnarray}
  I_{00}(k,q) &=& \sqrt{4 \pi}\,\int_0^{+\infty}
  dr\,r\,u(r)\,j_0(k r)\,j_0(\frac{q}{2} r) \, , \\
  I_{01}(k,q) &=& \frac{1}{\sqrt{4 \pi}}\,\int_0^{+\infty}
  dr\,u(r)\,j_0(k r)\,\frac{j_1(\frac{q}{2} r)}{(\frac{q}{2} r)} \, , \\
  I_{10}(k,q) &=& \frac{1}{\sqrt{4 \pi}}\,\int_0^{+\infty}
  dr\,u(r)\,\frac{j_1(k r)}{k r}\,j_0(\frac{q}{2} r) \, ,
\end{eqnarray}
with $j_n(x)$ the spherical Bessel functions.
We notice that in the theory we are using the reduced wave function
takes the form $u(r) = A_S\, e^{-\gamma r}$, for which the $I_{00}$, $I_{01}$ and
$I_{10}$ integrals can be evaluated analytically.

The scattering of the $DD$ in the final state can be taken into
account by changing the $j_0(k r)$ (which comes from assuming the
final $DD$ state is a plane wave) in the integral $I_{00}$ to
\begin{eqnarray}
  j_0(k r) \to \cos{\delta(k)}\,j_0(k r) - \sin{\delta(k)}\,y_0(k r) \, ,
\end{eqnarray}
where $\delta(k)$ is the S-wave $DD$ phase shift and $y_n$
the spherical Neumann functions. If we assume that scattering
in the final state is weak (as dictated by our counting),
we can simply approximate
\begin{eqnarray}
  \cos{\delta} \approx 1 \quad \mbox{and} \quad
  \sin{\delta} \approx - a_0 k \, ,
\end{eqnarray}
with $a_0$ the $DD$ scattering length~\footnote{Notice that in the sign
  convention we are using $k\,\cot{\delta(k)} \to -1/a_0$ for $k \to 0$,
  and a bound (virtual) state entails a positive (negative)
  scattering length.}.
Thus, the only change we have to do is the substitution
\begin{eqnarray}
  I_{00}(k, q) \to I_{00}(k,q) + (a_0 k)\,Y_{00}(k, q) \, ,
  \label{eq:DD-final}
\end{eqnarray}
with $Y_{00}$ defined as
\begin{eqnarray}
  Y_{00}(k,q) &=& \sqrt{4 \pi}\,\int_0^{\infty}
  dr\,r\,u(r)\,y_0(k r)\,j_0(\frac{q}{2} r) \, .
\end{eqnarray}
We notice that the combination of the $DD$ rescattering with the seagull
diagram, which is ${\rm N^2LO}$, is logarithmically divergent and
requires the inclusion of a contact-range two-body operator.
This represents a non-trivial check of our initial power counting
estimation for this operator.

For the decay of the $T_{cc}^+$ into $DD\gamma$ we use basically the same
formalism, though in this case there is no two-body operator:
the lowest order one enters at ${\rm N^2LO}$.
The decay amplitude takes the form
\begin{eqnarray}
  \langle DD(\vec{p}\,') \gamma | H | D^* D(\vec{p}\,) \rangle =
  A_{\rm M1}(\vec{p}\,',\vec{p}, \vec{q}) \, ,
\end{eqnarray}
with $A_{M1}$ given by
\begin{eqnarray}
  A_{M1} &=& \mu(D^*)\,\vec{\alpha} \cdot \left(
  \vec{\epsilon}_1 \times \vec{q} \right)\,
  (2 \pi)^3 \delta^{(3)} (\vec{p}\,' - \vec{p} + \frac{\vec{q}}{2})
  \, , \nonumber \\
\end{eqnarray}
where $\vec{\alpha}$ is the polarization vector of the photon and $\mu(D^*)$
the magnetic moment of the relevant $D^* \to D\gamma$ transition,
which if written in the isospin basis reads
\begin{eqnarray}
  \mu(D^*) = \mu_{+}\,\left( \frac{1+\tau_{1z}}{2} \right) +
  \mu_{0}\,\left( \frac{1-\tau_{1z}}{2} \right) \, ,
\end{eqnarray}
with $\mu_{+}$ and $\mu_0$ the magnetic moments for $D^{*+} \to D^+ \gamma$
and $D^{*0} \to D^0 \gamma$, though for this decay one might as well
simply use the particle basis.
The calculation of the decay width uses Eq.~(\ref{eq:fermi-golden-rule-DDpi})
but with the substitution $\omega \to q$ for adapting it to the photon case.
The matrix elements of $A_{M1}$ are obtained as before, leading to
\begin{eqnarray}
  \langle A_{M1} \rangle &=& \langle \mu(D^*) \rangle\,\vec{\alpha} \cdot \left(
  \vec{\epsilon}_1 \times \vec{q} \right)\,I_{00}(k, q) \, , 
\end{eqnarray}
which is completely analogous to Eq.~(\ref{eq:Ac-matrix-element}).
The inclusion of rescattering effects in the final $DD$ state is done
again with Eq.~(\ref{eq:DD-final}).

\section{Molecular $T_{cc}^+$ wave function}

%
If isospin symmetry were to be conserved in the masses,
the wave function of the $T_{cc}$ would be written as
\begin{eqnarray}
  \langle \vec{x}\,| T^{(I)}_{cc} \rangle &=& \psi(\vec{x})\,
  | D^* D (I) \rangle \, , 
\end{eqnarray}
depending on whether its isospin is $I=0$ or $1$ (where molecular models
show a clear preference for $I=0$).
However, the $T_{cc}^+$ is located merely a few hundred of ${\rm keV}$ below
the $D^{*+} D^0$, which is small in comparison with the mass difference
between the $D^{*+} D^0$ and $D^{*0} D^+$ thresholds
(about $1.4\,{\rm MeV}$).
For this reason we instead consider the $T_{cc}^+$ wave function to be a linear
combination of a {\it low} and a {\it high mass} channel contributions:
\begin{eqnarray}
  \langle \vec{x} | T_{cc} \rangle =
  \psi_L(\vec{x})\,| L \rangle +
  \psi_H(\vec{x})\,| H \rangle \, ,
\end{eqnarray}
with $| L \rangle$ and $| H \rangle$ given by
\begin{eqnarray}
  | L \rangle = | D^{*+} D^0 \rangle  \quad \mbox{and} \quad
  | H \rangle = | D^{*0} D^+ \rangle  \, .
\end{eqnarray}

For determining the wave function, we first consider the EFT expansion of
the $D^* D$ interaction
\begin{eqnarray}
  V_{\rm EFT}(\vec{q}) &=& C_I + D_I {\vec{q}\,}^2 \nonumber \\
  &+&  E_I\,\left( 
  \vec{\epsilon_1} \cdot \vec{q} \, \vec{\epsilon_1}^* \cdot \vec{q} -
  \frac{1}{3} \, {\vec{q}\,}^2 \right) \, \nonumber \\
  &+& V_{\rm OPE}(\vec{q}) \, ,
  \label{eq:V-SL}
\end{eqnarray}
where $\vec{q}$ is the exchanged momentum between the mesons, $I=0,1$ indicates
isospin, $C_I$ and $D_I$ represent momentum-independent and
momentum-dependent S-wave interactions,
$E_I$ is an S-to-D-wave contact interaction and $V_{\rm OPE}$ is
the one pion exchange (OPE) potential.
We will count $C_0$ as ${\rm LO}$, $C_1$ and $D_0$ as ${\rm NLO}$
and $E_0$ as ${\rm N^2LO}$.
As for the OPE potential it is nominally ${\rm NLO}$, but the actual
momentum scale at which central and tensor pion exchanges become
non-perturbative in the $D^*D$/$D\bar{D}^*$ systems has been estimated
to be $\Lambda_C > 1\,{\rm GeV}$ and $\Lambda_T = 290\,{\rm MeV}$
in~\cite{Valderrama:2012jv}, but this corresponds to $g_1 = 0.6$.
If we use the updated value of the axial coupling $g_1 = 0.56$,
the tensor scale will become $\Lambda_T = 330\,{\rm MeV}$.
The size of the tensor corrections is thus expected to be
$\gamma_L / \Lambda_T \sim 0.08$, which is in between
$\rm N^{1.7}LO$ and $\rm N^{2.9}LO$ for our estimation of
the expansion parameter ($(0.2)^{1.7}$ and $(0.4)^{2.9}$
are approximately $0.08$).
Yet, we warn that this estimation will require further attention:
owing to the effective mass of the pion being
relatively small ($\mu_{\pi} \to 0$),
the S-to-D wave tensor force effectively becomes a $1/r^3$ potential,
which has a really long range and might generate a D-wave component of
the wave function that is larger than expected.

As far as pions remain subleading, we have a contact theory
with a wave function of the type
\begin{eqnarray}
  \psi_L(\vec{x}) &=& \frac{A_S}{\sqrt{4 \pi}}\,\phi_L\,
  \frac{e^{-\gamma_L r}}{r} \, , \\
  \psi_H(\vec{x}) &=& \frac{A_S}{\sqrt{4 \pi}}\,\phi_H\,
  \frac{e^{-\gamma_H r}}{r} \, , 
\end{eqnarray}
with $A_S$ the asymptotic normalization of the wave function,
$\phi_L$ and $\phi_H$ (such that $|\phi_L|^2 + |\phi_H|^2 = 1$)
the amplitudes of the $L$ and $H$ channels and $\gamma_L = 26.4\,{\rm MeV}$
and $\gamma_H = 58.5\,{\rm MeV}$ the wave numbers
for the central value of $\delta m_{\rm pole}$.
It will prove useful to also define $\phi_L$ and $\phi_H$ in terms of
the isospin angle $\theta_I$:
\begin{eqnarray}
  \phi_L = \cos{\theta_I} \quad \mbox{and} \quad \phi_H = \sin{\theta_I} \, .
\end{eqnarray}
At ${\rm LO}$, if we assume that the $T_{cc}^+$ is predominantly an $I=0$ state
(at least at short distances), we will only have an isoscalar contact-range
interaction which basically fixes
$\phi^{\rm LO}_L = - \phi^{\rm LO}_H = 1/\sqrt{2}$ (modulo corrections
from the difference in the reduced masses of the $L$ and $H$ channels).
The ${\rm LO}$ asymptotic normalization will be determined by the normalization
of the wave function, i.e.
\begin{eqnarray}
  {|A^{\rm LO}_S|}^2 \, \int_0^{\infty} dr\,\left[ {|\phi^{\rm LO}_L|}^2 u_L^2(r) +
    {|\phi^{\rm LO}_H|}^2 u_H^2(r) \right] = 1 \, , \nonumber \\
\end{eqnarray}
with $u_L(r) = e^{-\gamma_L r}$ and $u_H(r) = e^{-\gamma_H r}$, from which
\begin{eqnarray}
  \frac{1}{A_S^{\rm LO}} &=&
  \sqrt{
    \frac{{\phi^{\rm LO}_L}^2}{2 \gamma_L} + \frac{{\phi^{\rm LO}_H}^2}{2 \gamma_H} 
  } \, . \label{eq:AS-LO}
\end{eqnarray}
At ${\rm NLO}$ we will have corrections from (i) the potential in the $I=1$
channel, i.e. $C_1$, which actually does not change the form of the
wave function at all (ii) the momentum-dependent contact-range
interaction in the $I=0$ channel, i.e. $D_0$, which breaks
the relation between $A_S$ and the normalization of
the wave function.
Basically this implies the correction:
\begin{eqnarray}
  A_S^{\rm NLO} = A_S^{\rm LO} + \delta A_S \, ,
\end{eqnarray}
Now two strategies are possible here: (i) to determine $\delta A_S$
from the isoscalar effective range or (ii) to determine
them from external information.
The first strategy is the more {usual} one in pionless EFT, though besides
expanding in terms of range corrections~\cite{Beane:2000fi}
it is also possible to expand in terms of
the wave function renormalization~\cite{Phillips:1999hh}.

The second strategy is equivalent (up to higher order corrections) to the first
one and might be easier to pull off simply because $A_S$ (and also $\theta_I$)
can be determined from potential models.
In fact, provided that the interaction binding the $T_{cc}^+$ is attractive
at most distance scales, we will have $A_S > A_S^{\rm LO}$ (the only way
in which to obtain $A_S < A_S^{\rm LO}$ is with an attractive short-range
potential surrounded by a repulsive barrier), which works
in the direction of increasing the decay widths.

Finally, for completeness and as a non-trivial crosscheck of isospin symmetry,
we will explicitly consider changes of the isospin angle at ${\rm NLO}$
\begin{eqnarray}
  \theta_I^{\rm NLO} = \theta_I^{\rm LO} + \delta \theta_I \, .
\end{eqnarray}
From the expansion of the potential in Eq.~(\ref{eq:V-SL}) we expect
$\theta_I^{\rm LO} \approx \theta_I^{\rm NLO} \approx -45\,{^\circ}$,
modulo negligible corrections from the difference in the reduced
masses of the $L$ and $H$ channel and the relative effect of
the range corrections (the $D_0$ coupling)
in these two channels.
Yet, considering a non-trivial $\delta \theta_I$ might reveal the existence
of isospin breaking contact terms at short distances (though here
short distances actually includes two-pion exchange diagrams
that might generate a larger than expected $\delta \theta_I)$.
We will see that the ${\rm NLO}$ calculation yields a $\theta_I^{\rm NLO}$
compatible with $-45^{\circ}$ when compared with the relevant
experimental data.

\section{Non-molecular component \\ of the $T_{cc}^+$}

%
Predictions for the $T_{cc}^+$ tetraquark fall into two categories depending
on whether they are based on its quark or charmed meson degrees of freedom.
We might loosely refer to them as compact and molecular.
This is not necessarily a clear-cut distinction though, as
four-quark explanations can perfectly generate a non-compact,
two-charmed meson component of the wave function if the mass of
the tetraquark happens to be close to the $D^*D$ threshold~\cite{Janc:2004qn}.
For the sake of simplicity we might consider that the $T_{cc}^+$ wave
function can be subdivided into a non-compact and compact component:
\begin{eqnarray}
  | T_{cc}^+ \rangle = \cos{\theta_C}\,| D^* D \rangle +
  \sin{\theta_C}\,| cc\bar{q}\bar{q} \rangle \, ,
\end{eqnarray}
with the non-compact piece corresponding to the $D^*D$ molecular explanation
we have referred to previously, while the compact piece represents
the non-molecular components.
$\theta_C$ represents the mixing angle between these two pieces of
the wave function, where $\theta_C = 0$ corresponds to the {usual}
molecular interpretation of the $T_{cc}^+$.

The interesting point is how a compact component will enter the description of
the decay widths.
The contribution of a wave function component to the decay amplitude depends
on the momentum scales involved.
For the molecular component we have
\begin{eqnarray}
  && \langle T_{cc} (D^* D) | A_c^{\rm 1B} | D D \pi \rangle = \nonumber \\
  && \quad \frac{g_1}{\sqrt{2} f_{\pi}}\,\vec{\epsilon}_1 \cdot \vec{q}\,
  \int d^3\vec{x}\,\langle \psi | \vec{x} \rangle
  e^{-i (\vec{k} - \vec{q}/2) \cdot \vec{x}} =
  \nonumber \\
  && \quad \frac{g_1}{\sqrt{2} f_{\pi}}\,\vec{\epsilon}_1 \cdot \vec{q}\,\,
  \langle \psi | \vec{k} - \vec{q}/2 \rangle \, ,
\end{eqnarray}
where $\langle \psi | \vec{x} \rangle$ and
$\langle \psi | \vec{p} \rangle$ are
the r- and p-space molecular wave functions.
By taking into account that in a contact-range theory
$\langle \psi | \vec{p} \rangle = \sqrt{8 \pi \gamma}/({p^2 + \gamma^2})$,
we expect that this matrix element scales as
\begin{eqnarray}
  && \langle T_{cc} (D^* D) | A_c^{\rm 1B} | D D \pi \rangle \propto
  \frac{g_1}{\sqrt{2} f_{\pi}}\,\,\vec{\epsilon}_1 \cdot \vec{q}\,\,
  \frac{\sqrt{2 \pi}}{Q^{3/2}} \, ,
\end{eqnarray}
with $Q \sim \gamma_L$, $\gamma_H$, $k$ or $q$ the characteristic low-energy
momentum scale for a molecular $T_{cc}^+$.
From this, the natural expectation for the scaling of the decay amplitude
of a compact component would be
\begin{eqnarray}
  && \langle T_{cc} (cc \bar{q}\bar{q}) | A_c | D D \pi \rangle \propto
  \frac{g_1'}{\sqrt{2} f_{\pi}}\,\,\vec{\epsilon}_1 \cdot \vec{q}\,
  \frac{\sqrt{2 \pi}}{M_{C}^{3/2}} \, ,
\end{eqnarray}
with $M_C$ the characteristic momentum scale for a compact tetraquark,
which we expect to be of the order of the natural hadronic scale
$M_C \sim (0.5-1.0)\,{\rm GeV}$ (and $g_1'$ the coupling of
the compact component to $D D \pi$, which we have assumed to
be of roughly the same size as $g_1$).
This scaling argument also applies to the $DD\gamma$ decays.

From the previous the decay amplitude of a compact component of
the $T_{cc}$ wave function is expected to be suppressed by
a factor of $(Q/M_C)^{3/2}$ with respect to
the ${\rm LO}$ contribution.
Were $M_C$ to be of the order of the hard scale in the EFT we are using here,
i.e. $M \sim (1-2) m_{\pi}$, the contribution from a compact component
to the decay would enter at ${\rm N^{3/2} LO}$.
It turns out that $M_C > M$, which means that this contribution
enters at a considerably higher order.
Thus, at lower orders in the EFT expansion, the effect of a compact
tetraquark component is simply to reduce the total decay width:
\begin{eqnarray}
  \Gamma(T_{cc}^+) = \cos^2{\theta_C}\,\Gamma(T_{cc}^+(D^*D)) \, .
\end{eqnarray}
That is, if a molecular prediction overshots the experimental decay width
by a noticeable amount, this might indicate the existence of
a non-molecular component for the $T_{cc}^+$ tetraquark.

It is however worth noticing that the explicit separation of the $T_{cc}^+$
wave function into molecular and non-molecular components generates a
parameter redundancy problem, as the observable effects of the compact
mixing angle $\theta_C$ can be reabsorbed into the EFT's subleading
range corrections, i.e. into $A_S$.
Indeed, at ${\rm NLO}$
the decay amplitude is proportional to these two factors
\begin{eqnarray}
  \Gamma(T_{cc}^+) \propto \cos^2{\theta_C}\,A_S^2 \, ,
\end{eqnarray}
which means that {\it compactness} can be recast into a negative contribution
to the effective range (as this reduces $A_S^2$, see Appendix~\ref{app:AS}).
Thus, the angle $\theta_C$ should be considered as a model-dependent quantity,
at least in the absence of a model-independent disentanglement of the dynamics
between the molecular and non-molecular degrees of freedom.
Unfortunately, though the inclusion of a compact $T_{cc}^+$ field
is straightforward, this still does not resolve
the parameter redundancy problem
(which probably requires invoking phenomenological models).

\section{Couplings}

%
The width of a molecular $T_{cc}$ depends on the axial coupling $g_1$ and
the magnetic moments $\mu_+$ and $\mu_0$ for the $D^*$ to $D$ transitions,
which can be extracted from the decay widths of the charmed mesons.
We begin with $g_1$, for which we use the decays of $D^{*+}$ into $D \pi$
\begin{eqnarray}
  \Gamma (D^{*+} \to D^0 \pi^+) &=& \frac{g_1^2}{6 \pi f_{\pi}^2}\,
  \frac{m_{D^0}}{m_{D^{*+}}}\,q_{\pi^+}^3 \, , \\
  \Gamma (D^{*+} \to D^+ \pi^0) &=& \frac{g_1^2}{12 \pi f_{\pi}^2}\,
    \frac{m_{D^+}}{m_{D^{*+}}}\,q_{\pi^0}^3 \, , 
\end{eqnarray}
where $f_{\pi} = 130\,{\rm MeV}$ and $q_{\pi}$ the momentum of the emitted pion.
From the $D^{*+}$ decay width and branching ratios provided in
the Review of Particle Physics (RPP)~\cite{Zyla:2020zbs}, i.e.
$\Gamma(D^{*+}) = 83.4 \pm 1.8 \,{\rm keV}$,
$\Gamma(D^0 \pi^+) / \Gamma = (67.7\pm 0.5)\%$ and
$\Gamma(D^+ \pi^0) / \Gamma = (30.7\pm 0.5)\%$,
we obtain $g_1 = 0.56 \pm 0.01$.

For the magnetic moments $\mu_{+}$ and $\mu_0$ we use the $D^*$ decays
into $D \gamma$
\begin{eqnarray}
  \Gamma (D^{*} \to D \gamma) &=& \frac{|\mu|^2}{3 \pi}\,
  \frac{m_D}{m_{D^*}}\,q_{\gamma}^3 \, ,
\end{eqnarray}
with $q_{\gamma}$ the momentum of the outgoing photon.
For $\mu_+$ we use again the $D^{*+}$ decay width and its branching ratio
into $D^+ \gamma$ (i.e. $1.6 \pm 0.4\%$), yielding
$\mu_+ = 0.46 \pm 0.06\,\mu_{n.m.}$ where the sign is chosen as to
coincide with that of the magnetic moment of the $\bar{d}$
antiquark within the $D^{*+}$ and with $\mu_{n.m.} = |e| / 2 m_N$
the nuclear magneton.
The determination of $\mu_0$ is more indirect as the $D^{*0}$ decay width is
not experimentally known (beyond an upper bound).
However its branching ratios into $D^0 \pi^0$ and $D^0 \gamma$ are well
determined~\cite{Zyla:2020zbs} and the partial decay width
into $D^0 \pi^0$ can be calculated from $g_1$ (resulting in
$\Gamma (D^{*0} \to D^0\pi^0) = 35.9 \pm 1.3\,{\rm keV}$),
which all together yields
$\Gamma (D^{*0} \to D^0\gamma) = 19.6 \pm 1.0\,{\rm keV}$.
From this we obtain $\mu_0 = -(1.72\pm 0.05)\,\mu_{n.m.}$.

\section{Partial decay widths}

%
With the previous ingredients we are ready to calculate the $T_{cc} \to DD \pi$
and $T_{cc} \to DD \gamma$ decay widths.
For this we have to sandwich the decay operator between the initial and
final states, which though laborious (we have to take into account
isospin breaking in the $L$ and $H$ channels)
it is nonetheless straightforward.
We will use the $\delta m_{\rm pole}$ solution, from which the binding
energy of the $L$ ($H$) components of a molecular $T_{cc}^+$ is
$B_L = 0.36 \pm 0.04 \,{\rm MeV}$ ($B_H = 1.77 \pm 0.04\,{\rm MeV}$).
In addition, we will assume a purely molecular $T_{cc}^+$ (i.e.
$\theta_C = 0$) unless stated otherwise.

We begin with the tree level amplitudes.
For the $T_{cc}^+ \to D^0 D^0 \pi^+$ decay, the final state contains
two identical bosons and requires symmetrization, which is done
by adding the $A(\vec{k},\vec{q})$ and $A(-{\vec{k},\vec{q}})$
amplitudes and then changing the phase space factor
for the final $D^0D^0$ pair from $d^3{\vec{k}} \to d^3{\vec{k}}/2$ to avoid
counting the final $D^0 D^0$ states twice.
We obtain
\begin{eqnarray}
  \Gamma^{\rm LO (1B)}(T_{cc}^+ \to D^0 D^0 \pi^+) =
  29.6^{+1.1}_{-1.0} \, {}^{+1.8}_{-1.8}\,{\rm keV} \, , \\
  \Gamma^{\rm LO (1B)}(T_{cc}^+ \to D^+ D^0 \pi^0) =
  13.7^{+0.5}_{-0.5} \, {}^{+0.5}_{-0.6}\,{\rm keV} \, , \\
  \Gamma^{\rm LO (1B)}(T_{cc}^+ \to D^+ D^0 \gamma) =
  5.8 \pm 0.4 \, \pm 0.2\,{\rm keV} \, , 
\end{eqnarray}
which basically agrees with~\cite{Meng:2021jnw} and
where we have taken $\phi_L = - \phi_H = 1/\sqrt{2}$ and
$A_S = 8.5\,{\rm MeV}^{1/2}$ (obtained from the normalization of
the wave function).
The first uncertainty corresponds to varying $g_1$ for the strong decays
and $\mu_+$ and $\mu_0$ for the electromagnetic one, while the second
comes from the binding energy.
We notice that the previous amplitudes only takes into account a final $D^0 D^0$
state in S-wave or a final $D^+ D^0$ state in S- or P-wave: adding
the contributions from higher $L=2,4,6,...$ ($L=2,3,4,...$)
partial waves of the $D^0 D^0$ ($D^+ D^0$) final state
will change the partial decay widths to $29.9$,
$14.1$ and $6.4\,{\rm keV}$, respectively,
i.e. a small $1.3\,{\rm keV}$ increase
in the total decay width.
With the exception of the electromagnetic decay, the decay widths
increase if the binding energy is reduced (e.g. if we use
$\delta m_{\rm BW}$ instead of $\delta m_{\rm pole}$ we would
obtain $33.7$, $14.9$ and $5.5\,{\rm keV}$
for the partial decay widths).
The combined $\rm LO$ decay width is
\begin{eqnarray}
  \Gamma^{\rm LO}(T_{cc}^+) &=&
  49.1^{+1.6}_{-1.5}\,{}^{+0.4}_{-0.4}\,{}^{+2.2}_{-2.1}\,{\rm keV} \, , \nonumber \\
  &=& 49.1^{+2.7}_{-2.6}\,{\rm keV} \, ,
\end{eqnarray}
where the uncertainties refer only to the input parameters ($g_1$,
$\mu_{+}$/$\mu_0$ and $\delta m$), not to the EFT
convergence rate (which we have not discussed yet).
{
  It is interesting to notice that this width is in line with most of
  the other $\rm LO$ calculations available: $47\,{\rm keV}$
  in~\cite{Meng:2021jnw} (which uses $\delta m_{\rm BW}$
  instead of $\delta m_{\rm pole}$), $53\,{\rm keV}$ in
  \cite{Ling:2021bir} (which calculates the decay width
  in the isospin limit), $43\,{\rm keV}$ ($80\,{\rm keV}$)
  for $\delta m_{\rm pole}$ ($\delta m_{\rm BW}$) in~\cite{Feijoo:2021ppq}
  (which directly convolutes the width of the charmed mesons to obtain
  the $T_{cc}^+$ width) and $52\,{\rm keV}$ in~\cite{Fleming:2021wmk}
  (which also uses $\delta m_{\rm BW}$).
  A cursory comparison of the previous predictions suggest a ${\rm LO}$
  error of the order of $10\,{\rm keV}$, a figure 
  compatible with the EFT uncertainties we will later obtain
  in Eqs.~(\ref{eq:EFT-conv}) and (\ref{eq:LO-w-errors}).
  }

Next we consider the rescattering of the $D D$ pair in the final state,
which requires the scattering length of this system as input.
The only phenomenological calculation of this quantity we are aware of
is Ref.~\cite{Liu:2019stu}, which estimates
$a_0(DD) = -0.4^{+0.1}_{-0.2}\,{\rm fm}$ (and also predicts
$\delta m = -3^{+4}_{-15}\,{\rm MeV}$ for the $T_{cc}$ in the isospin symmetric
limit, from which we may assume that the actual $DD$ scattering length
will also fall within the error bars~\footnote{We mention though that
  calculations of the two-bottom-meson potential in the lattice
  indicate that the $I=1$ $B B$ configuration is attractive
  overall~\cite{Detmold:2007wk} (which in our sign convention
  will generate $a_0 < 0$ if the attraction is not strong enough
  as to generate a bound state), particularly at short distances.
  Chiral EFT also predicts an
  attractive two-pion exchange potential for $I=1$ $B B$~\cite{Wang:2018atz}.
  Finally, from heavy flavor symmetry we expect the $D D$ and $B B$
  potentials to be similar.}).
If we use the ${\rm LO}$ values of $A_S$, $\phi_L$ and $\phi_H$ (which we will
from now on, unless stated otherwise), we obtain
\begin{eqnarray}
  \Gamma^{\rm (1B + DD)}(T_{cc}^+ \to D^0 D^0 \pi^+) &=&
  33.3^{+1.2}_{-1.2}\,{}^{+1.9}_{-1.7}\,{}^{+2.0}_{-0.9}\,
  {\rm keV} \, , \nonumber \\ \\
  \Gamma^{\rm (1B + DD)}(T_{cc}^+ \to D^+ D^0 \pi^0) &=&
  15.9^{+0.6}_{-0.5}\,{}^{+0.5}_{-0.5}\,{}^{+1.2}_{-0.6}\,
  {\rm keV} \, , \nonumber \\ \\
  \Gamma^{\rm (1B + DD)}(T_{cc}^+ \to D^+ D^0 \gamma) &=&
  7.5 \pm 0.6\,\pm 0.2\,{}^{+0.9}_{-0.4}\,
    {\rm keV} \, , \nonumber \\
\end{eqnarray}
where the source of the first two errors is as in the ${\rm LO}$ calculation
and the third error comes from the propagation of the uncertainty in $a_0$.
{We stress that in the counting used here the $DD$ interaction is
  perturbative.
  Previously, a non-perturbative final state interaction has been considered
  for instance in case of the $X(3872)$ as a $D^*\bar{D}$ system and
  its decays into $D\bar{D} \pi$~\cite{Guo:2014hqa,Dai:2019hrf}.
  For the $T_{cc}^+$, Ref.~\cite{Fleming:2021wmk}
  has recently considered the case in which the final $DD$ interaction
  is able to form a bound state.
}

Then we consider the inclusion of the seagull diagram (but without including
the $DD$ rescattering or the changes in asymptotic normalization),
which only affects the $DD\pi$ decays, arriving at
\begin{eqnarray}
  \Gamma^{\rm (1B + 2B)}(T_{cc}^+ \to D^0 D^0 \pi^+) &=&
  30.2^{+1.1}_{-1.0}\,{}^{+1.8}_{-1.6}\,{\rm keV} \, , \\
  \Gamma^{\rm (1B + 2B)}(T_{cc}^+ \to D^+ D^0 \pi^0) &=&
  14.1^{+0.6}_{-0.4}\,{}^{+0.7}_{-0.5}\,{\rm keV} \, , 
\end{eqnarray}
which implies that the two-body corrections are actually smaller than
the rescattering of the final $D D$ mesons and where the uncertainties
are the same as in the ${\rm LO}$ calculation
($g_1$ and $\delta m_{\rm pole}$).
{
  Here a comparison with the $D^*\bar{D}$ system --- the $X(3872)$ ---
  is in order: Ref.~\cite{Dai:2019hrf}, which previously considered
  the rescattering and two-body corrections for the $X(3872)$ decays,
  also arrived to the conclusion that rescattering effects
  are much larger than the seagull diagram, where the later
  contribution happens to be fairly small.
}

Finally, including these two subleading order corrections together
(but using the ${\rm LO}$ values of the wave function parameters),
we obtain an {\it abridged} ${\rm NLO}$ result for the decay widths:
\begin{eqnarray}
  \Gamma^{\rm NLO(*)}(T_{cc}^+ \to D^0 D^0 \pi^+) &=&
  34.0 \pm 1.2 \,{}^{+1.8}_{-1.7} \, {}^{+1.9}_{-1.0}
  \,{\rm keV} \, , \nonumber \\ \\
  \Gamma^{\rm NLO(*)}(T_{cc}^+ \to D^+ D^0 \pi^0) &=&
  16.4^{+0.6}_{-0.5}\,{}^{+0.6}_{-0.5}\,{}^{+1.2}_{-0.5}
  \,{\rm keV} \, , \nonumber \\ \\
  \Gamma^{\rm NLO(*)}(T_{cc}^+ \to D^+ D^0 \gamma) &=&
  7.5\pm 0.6\,\pm 0.2\,{}^{+0.9}_{-0.4}\,
    {\rm keV} \, , \nonumber \\
\end{eqnarray}
where the uncertainties are as in the previous $DD$ rescattering partial widths
($g_1$ or $\mu_{+}$ and $\mu_0$, $\delta m_{\rm pole}$ and $a_0$).
The combined decay width will then be
\begin{eqnarray}
  \Gamma^{\rm NLO(*)}(T_{cc}^+) &=&
  57.9^{+1.8}_{-1.8}\,\pm 0.6\,{}^{+4.1}_{-2.0}\,{}^{+2.2}_{-2.1}\,{\rm keV} 
  \nonumber \\
  &=& 57.9^{+5.0}_{-3.4} \,{\rm keV} \, , 
\end{eqnarray}
where for the moment the uncertainty only refer to the one coming
from the input parameters of the calculation.
If we are interested in the EFT uncertainty, we can compare the abridged
$\rm NLO(*)$ decay width with the $\rm LO$ one, suggesting 
\begin{eqnarray}
  \left| \frac{\Gamma^{\rm NLO(*)} - \Gamma^{\rm LO}}{\Gamma^{\rm LO}}
  \right| \approx 0.18 \, ,
\end{eqnarray}
in line with the naive estimations in Eq.~(\ref{eq:naive-EFT-convergence})
of a convergence rate within $0.2-0.4$.
However, this does not take into account the possible corrections to the
asymptotic normalization and isospin angle, which might worsen
the convergence of the EFT.
For exploring what to expect from the corrections to the asymptotic
normalization, we might look at the two nucleon system.
There the ${\rm LO}$ wave function in a pionless theory would
yield $A^{\rm LO}_S = \sqrt{2 \gamma_d} = 0.6806\,{\rm fm}^{-1/2}$,
which is to be compared
with $A_S = 0.8846(9)\,{\rm fm}^{-1/2}$~\cite{deSwart:1995ui},
yielding $A_S^2 / {(A^{\rm LO}_S)}^2 = 1.69$.
Were this ratio to hold for the $T_{cc}^+$ case, we would have a
$70\%$ increase in the ${\rm NLO}$ decay width if we were
to expand directly in terms of $A_S$.
This is probably not the case, though: the $A_S^2 / {(A^{\rm LO}_S)}^2$ ratio
scales as $1 / (1 - \gamma r_e)$ with $\gamma$ the binding momentum and
$r_e$ the effective range (with this simple approximation resulting
in $1.68$ for the deuteron).
The $T_{cc}^+$ is less bound than the deuteron though and naively
we expect range corrections in the $T_{cc}^+$ to be of
the same order of magnitude as those of the $X(3872)$,
which are considerably smaller than
in the deuteron case~\cite{Fleming:2007rp,Dai:2019hrf}.
If we notice that non-tensor OPE almost cancels in the $D^*D$ and
$D^*\bar{D}$ systems, the scale of range corrections is probably
set by the tensor scale $\Lambda_T = 330\,{\rm MeV}$
(check the previous discussion below Eq.(\ref{eq:V-SL})),
while in the deuteron the range will be given by
the pion mass, indicating that the expected range of the $D^* D$ potential
is about $0.42$ times that of the two-nucleon case.
From this we could expect $A_S^2/ {(A^{\rm LO}_S)}^2 \sim 1.1$, which will
suggest a convergence parameter of
\begin{eqnarray}
  \left| \frac{\Gamma^{\rm NLO} - \Gamma^{\rm LO}}{\Gamma^{\rm LO}}
  \right| \approx 0.3 \, . \label{eq:EFT-conv}
\end{eqnarray}
If this estimation were to hold, the full uncertainties in the ${\rm LO}$
and ${\rm NLO}$ calculation would be
\begin{eqnarray}
  \Gamma^{\rm LO}(T_{cc}^+) &=& 49 \pm 3 \pm 16 \,{\rm keV} \, ,
  \label{eq:LO-w-errors} \\
  \Gamma^{\rm NLO(*)}(T_{cc}^+) &=& 58^{+5}_{-3} \pm 5 \,{\rm keV} \, ,
  \label{eq:NLO-w-errors}
\end{eqnarray}
where the first and second errors refer to the input parameters and
the intrinsic EFT uncertainty, respectively.
But again, there might be factors which we have not properly considered
and which might alter the current conclusions, which should be taken
as temporary.
{As a crosscheck, we notice that most ${\rm LO}$ calculations
  available~\cite{Meng:2021jnw,Ling:2021bir,Feijoo:2021ppq,Fleming:2021wmk}
  lie within the $(43-53)\,{\rm keV}$ window and
  are thus compatible with our ${\rm LO}$ result
  within EFT uncertainties.}

Of course, the previous ${\rm NLO}$ partial decay widths are incomplete:
the full ${\rm NLO}$ calculation requires a modification
in the asymptotic normalization $A_S$ and the isospin
angle $\theta_I$, and in principle it is also possible
to consider the mixing angle between a non-compact and
compact $T_{cc}^+$ component, $\theta_C$.
Luckily these contributions can be factored out easily,
leading to the following expressions
\begin{eqnarray}
  && \Gamma^{\rm NLO}(T_{cc}^+ \to D^0 D^0 \pi^+) = \cos^2{\theta_C}\,A_S^2
  \,\times\,\Big(
  \nonumber \\
  && \quad \phi_L^2
  \left[ 0.825(94) - 0.248(17)\,a_0 + 0.0188(7)\, a_0^2 \right] +
  \nonumber \\
  && \quad \phi_L\,\phi_H
  \left[ -0.00648(47) + 0.000947(35)\,a_0 \right ] \,\Big) \, ,
  \nonumber \\ \\
  && \Gamma^{\rm NLO}(T_{cc}^+ \to D^+ D^0 \pi^0) = \cos^2{\theta_C}\,A_S^2
  \,\times\,\Big(
  \nonumber \\
  && \quad \phi_L^2
  \left[ 0.187(22) - 0.0552(39)\,a_0 + 0.00412(15)\,a_0^2 \right] +
  \nonumber \\
  && \quad \phi_L\,\phi_H
  \left[-0.164(13) + 0.0729(48)\,a_0 - 0.00719(27)\,a_0^2 \right ] +
  \nonumber \\
  && \quad \phi_H^2
  \left[
    0.0386(23) - 0.00230(11)\,a_0 + 0.00347(14)\,a_0^2 \right ] \,\Big) \, ,
  \nonumber \\ \\
  && \Gamma^{\rm NLO}(T_{cc}^+ \to D^+ D^0 \gamma) =  \cos^2{\theta_C}\,A_S^2
  \,\times\,\Big(
  \nonumber \\
  && \quad \phi_L^2
  \left[ 0.0119(30) - 0.0057(14)\,a_0 + 0.00078(19)\,a_0^2 \right] +
  \nonumber \\
  && \quad \phi_L\,\phi_H
  \left[-0.0624(82) + 0.0381(49)\,a_0 - 0.000573(73)\,a_0^2 \right ] +
  \nonumber \\
  && \quad \phi_H^2
  \left[ 0.0851(45) - 0.0638(34)\,a_0 + 0.0122(6)\,a_0^2 \right ] \,\Big) \, ,
  \nonumber \\
\end{eqnarray}
which return the partial decay widths in ${\rm keV}$ and require
as input $A_S$ in units of ${\rm MeV}^{-1/2}$ and $a_0$
in units of ${\rm fm}$.
The uncertainties are shown in parentheses and correspond to $g_1$,
$\mu_+$, $\mu_0$ and $\delta m_{\rm pole}$ summed in quadrature and
symmetrized (as these errors are almost symmetrical).
The term proportional to $\phi_H^2$ in $\Gamma(T_c^{+} \to D^ 0D^0 \pi^+)$
is actually negligible (a small contribution coming from the seagull diagram)
and we have not written it down.
The same could be argued of a few of the terms we have kept,
but in these cases the difference is at most of two orders of
magnitude, potentially up to about a $1\%$
difference for $|a_0| = 1\,{\rm fm}$.
Here it is worth reminding that these formulas are only expected to be valid
for $m_{\pi} a_0 < 1$, i.e. $a_0 < 1.4\,{\rm fm}$ (otherwise a different power
counting in which the interaction of the final $DD$ pair is non perturbative
should be used).
The actual accuracy of these formulas is limited by the EFT convergence,
where for our estimation of $0.3$ for the expansion parameter
we should expect the previous expressions to have
a $9\%$ uncertainty.

\section{Comparison with experiment}

%
The theoretical decay widths can be compared with the experimental analysis
to obtain information about the $T_{cc}^+$.
If we begin with the ${\rm LO}$ calculation, we quickly realize that
the EFT calculation overshots the experimental decay width as
extracted from the unitarized Breit-Wigner profile~\cite{LHCb:2021auc}
\begin{eqnarray}
  \frac{\Gamma_{\rm pole}}{\Gamma_{\rm LO}} =
  0.98^{+0.04}_{-0.22}\,{}^{+0.06}_{-0.06} {}^{+0.42}_{-0.23} \, ,
\end{eqnarray}
where the first uncertainty comes from $\Gamma_{\rm pole}$ while the second
and third are derived from the input parameters and the convergence rate
of $\Gamma_{\rm LO}$, respectively.
The previous figure is compatible with $1$.
If we allow for a non-trivial $\theta_C$, this ratio could be related to
the molecular content of the $T_{cc}^+$
\begin{eqnarray}
  \frac{\Gamma_{\rm pole}}{\Gamma_{\rm LO}} = \cos^2{\theta^{\rm LO}_C} \, ,
\end{eqnarray}
yielding
\begin{eqnarray}
  |\theta_{\rm C}^{\rm LO}| = (8.6_{-8.6}^{+7.7}\,{}^{+21.3}_{-8.6})^{\circ} \, ,
\end{eqnarray}
which is compatible with the absence of a compact component,
i.e. $\theta_C = 0$, within errors.

Other interesting experimental information in~\cite{LHCb:2021auc}
is the signal yields of the $T_{cc}^+$ to the $D^+ D^0$ and $D^0 D^0$
channels:
\begin{eqnarray}
  N_S^+ &\equiv& N_S(D^+ D^0) = 171\pm 26 \, , \\
  N_S^0 &\equiv& N_S(D^0 D^0) = 263 \pm 23 \, .
\end{eqnarray}
As the number of events grow, the ratio of these two numbers is expected to
approach the ratio of the decays to the $D^+ D^0$ and $D^0 D^0$ channels
\begin{eqnarray}
  \frac{\Gamma(T_{cc}^+ \to D^+ D^0 \pi^0/\gamma)}{
    \Gamma(T_{cc}^+ \to D^0 D^0 \pi^+)} &=&
  \frac{N_S^+}{N_S^0}
  \left( 1 + \mathcal{O}\left(
    \frac{1}{\sqrt{N_S}} \right) \right) \, , \nonumber \\
\end{eqnarray}
with $N_S = N_S^+ + N_S^0$.
Actually the error from the finite number of signals can be easily estimated
by assuming a binomial distribution for $N_S^+$ and $N_S^0$ and
finding the expected $68\%$ band for the $N_S^+ / N_S^0$ ratio.
Putting the pieces together, we arrive at
\begin{eqnarray}
  \frac{\Gamma^+}{\Gamma^0} = 0.65 \pm 0.10 {}^{+0.6}_{-0.5} {}^{+0.7}_{-0.6} =
  0.65^{+0.14}_{-0.13} \, ,
\end{eqnarray}
where the first two uncertainties come from $N_S^+$ and $N_S^0$, the last
one from the finite size of $N_S^+ + N_S^0$ and
then we add them in quadrature.
The LO ratio is
\begin{eqnarray}
  \frac{\Gamma^+_{\rm LO}}{\Gamma^0_{\rm LO}} = 0.66^{+0.03}_{-0.03}\,\pm 0.20 \, ,
\end{eqnarray}
which is compatible with the experimental yields.

At ${\rm NLO}$ the $\Gamma^+ / \Gamma^0$ ratio depends on the isospin angle,
where we find that reproducing the experimental ratio requires:
\begin{eqnarray}
  \theta_I^{\rm NLO} =
  (- 42.4^{+8.1}_{-6.1}\,{}^{+1.7}_{-1.6}\,\pm 3.8) \,{}^{\circ} \, ,
\end{eqnarray}
where the first uncertainty is experimental, the second are
the couplings / $a_0$ / $\delta m_{\rm pole}$ and the third
the ${\rm NLO}$ uncertainty.
This in turn implies that
\begin{eqnarray}
  \frac{\Gamma_{\rm pole}}{\Gamma_{\rm NLO}} =
  \cos^2{\theta_{\rm C}}\,\frac{A_S^2}{A_{S(C)}^2} =
  0.81^{+0.03}_{-0.21}\,{}^{+0.04}_{-0.06}\,\pm 0.07 \, , 
\end{eqnarray}
where the errors are as before and $A_{S(C)}$ refers to the normalization
for a contact-range theory with the isospin angle $\theta_I^{\rm NLO}$,
i.e. Eq.(\ref{eq:AS-LO}) but using $\theta_I^{\rm NLO}$ instead of
$\theta_I^{\rm LO} = -45\,{}^{\circ}$.
If we assume that $A_S = A_{S(C)}$ and the discrepancy comes exclusively
from the probability of the compact component, we will obtain
\begin{eqnarray}
  |\theta_{\rm C}^{\rm NLO}| =
  (25.7^{+13.0}_{-2.3} \,{}^{+4.2}_{-2.3} \,\pm 2.3) {}^{\circ} \, ,
\end{eqnarray}
with the errors as before and which would imply a $\theta_C$
distinguishable from zero at ${\rm NLO}$.
However, this is contingent to two factors: what would be $A_S$ for
the molecular component (if considered as a separate degree of freedom
from the compact one) and the fact that $\theta_C$ is not model
independent in the sense that its effects can be recast
into a negative effective range instead (i.e. the compact
component can be reabsorbed as energy dependence
in the molecular one).
In the first case, assuming $A_S^2/A_{S(C)}^2 = 1.1 (1.2)$ will
entail a compact mixing angle of $30.7 (34.6)\,{}^{\circ}$,
larger than the one we have calculated.
In the second case, we set $\theta_C = 0$ and recast the effects of
$\theta_C \neq 0$ into a negative effective range by means of
the formula (check Appendix~\ref{app:AS})
\begin{eqnarray}
  \frac{A_S^2}{A_{S(C)}^2} = \frac{1}
       {1 - \frac{1}{2}\,r_{e0}\,\left(\frac{1-\sin{2\theta_I}}{2}\right)
         \,A_{S(C)}^2} \, ,
\end{eqnarray}
where $r_{e0}$ is the $I=0$ effective range, leading to
\begin{eqnarray}
  r_{e0}^{\rm NLO}
  = -1.3^{+0.3}_{-2.3}\,{}^{+0.3}_{-0.6}\,\pm 0.1\,{\rm fm} \, ,  
\end{eqnarray}
which is, as expected, negative and within the confidence
limits (CL) of Ref.~\cite{LHCb:2021auc}
(i.e. $0 \geq r_{eL} \geq -11.9(-16.9) \, {\rm fm}$ within $90(95)\%$ CL,
where $r_{eL}$ refers to the effective range in the L channel;
if we assume no interaction in the isovector channel,
we will have $r_{eL} = 2\,r_{e0} - 1/\sqrt{\gamma_H^2 - \gamma_L^2}$, or
$r_{eL} = -6.4\,{\rm fm}$ for $r_{e0} = -1.3\,{\rm fm}$).

Alternatively, had we simply assumed $\theta_I = -45\,{}^{\circ}$,
the $\Gamma^+ / \Gamma_0$ ratio would have been
\begin{eqnarray}
  \frac{\Gamma^+_{\rm NLO}}{\Gamma^0_{\rm NLO}} \Big|_{\theta_I = -\frac{\pi}{4}} =
  0.70 \pm 0.03 \pm 0.06 \, ,
\end{eqnarray}
which is still compatible with the experimental yields.
Meanwhile, the non-molecular ratio would have been
\begin{eqnarray}
  \frac{\Gamma_{\rm pole}}{\Gamma_{\rm NLO}} \Big|_{\theta_I = -\frac{\pi}{4}} =
  \cos^2{\theta_{\rm C}}\,\frac{A_S^2}{A_{S(LO)}^2} =
  0.83^{+0.03}_{-0.21}\,{}^{+0.05}_{-0.06}\,\pm 0.07 \, ,
  \nonumber \\
\end{eqnarray}
which also happens to be different from $1$, again.
Assuming $A_S = A_{S(\rm LO)}$, this ratio would in turn imply a compact
mixing angle of
\begin{eqnarray}
  |\theta_{\rm C}^{\rm NLO}| \Big|_{\theta_I = -\frac{\pi}{4}} =
  (24.1^{+13.5}_{-2.3} \,{}^{+4.2}_{-2.3} \,\pm 2.3) {}^{\circ} \, ,
\end{eqnarray}
or, alternatively, assuming that the ratio comes exclusively from range
corrections, would imply an isoscalar effective range of
\begin{eqnarray}
  r_{e0}^{\rm NLO} \Big|_{\theta_I = -\frac{\pi}{4}} =
  -1.1^{+0.3}_{-2.2}\,{}^{+0.4}_{-0.5}\,\pm 0.1\,{\rm fm} \, , \nonumber \\ 
\end{eqnarray}
which is negative.
As can be appreciated, all the numbers obtained for $\theta_I = -45^{\circ}$
are indistinguishable within errors to the ones we obtain from fixing
$\theta_I$ to the $\Gamma_{+} / \Gamma_0$ ratio derived
from the experimental yields.

\section{Summary}

%
The $T_{cc}^+$ represents not only a fascinating discovery but also a
wonderful opportunity for the study of hadron spectroscopy and decays.

While we do not know for sure its nature yet --- the mass of the $T_{cc}^+$ is
in principle compatible with previous predictions of $I=0$, $J=1$
$cc \bar{u}\bar{d}$ compact tetraquarks and $D^* D$ shallow
bound states, its closeness to the $D^{*+} D^0$ threshold indicates
that at least part of its wave function will be $D^* D$.
This last component is amenable to relatively straightforward theoretical
treatments, including the calculation of its expected width, which
would be a crucial piece of information if we want to eventually
know the structure of the $T_{cc}^+$.

The physical scales involved in the molecular components of
the $T_{cc}^+$ indicate a moderate convergence rate
for the calculation of its decay width and a ${\rm NLO}$
calculation is required to achieve an accuracy comparable
with the $25$ to $40\%$ relative uncertainty
in the experimental result (depending
on the resonance profile used).
Our preliminary calculation shows that the inclusion of a seagull pion
decay operator and the rescattering of the final $DD$ pair increase
the total decay width of the $T_{cc}^+$ state from
$50\,{\rm keV}$ in ${\rm LO}$ to
$58\,{\rm keV}$ in our {\it abridged} ${\rm NLO}$ calculation
(i.e. a ${\rm NLO}$ calculation with a ${\rm LO}$ wave function).
This is still preliminary: there are corrections coming from
the asymptotic normalization $A_S$ of a molecular $T_{cc}^+$ state,
the particular isospin mixing $\theta_I$ between its $D^{*+} D^0$
and $D^{*0} D^+$ components and the final physical pion
rescattering with the charmed mesons at non-zero energy
(not to mention that the $DD$ system might interact more strongly
than we expect).
$A_S$ and $\theta_I$ could be easily estimated from phenomenological models
and fed into the ${\rm NLO}$ calculation, where we expect a moderate
increase from the $58\,{\rm keV}$ figure we obtain.

If we compare the ${\rm NLO}$ results with the total decay width extracted
from the unitarized Breit-Wigner profile
($\Gamma_{\rm pole} = 48^{+2}_{-12}\,{\rm keV}$),
the previously discussed factors (particularly $A_S$) point towards
an excess decay width for a purely molecular explanation of
the $T_{cc}^+$ at ${\rm NLO}$.
This excess can be interpreted as the existence of a compact component,
where the ratio of $\Gamma_{\rm pole}$ and $\Gamma_{\rm NLO}$ suggest
that the non-molecular probability of the $T_{cc}^+$ wave
function is about $20\%$.
This conclusion should be taken as temporary though, as future experimental
refinements regarding the $T_{cc}^+$ mass and decay widths might alter
the present picture.
In addition, the interplay of the compact and molecular components of
the $T_{cc}^+$ could also be improved.
Nonetheless, we find it worth mentioning that the picture that emerges
from the current EFT description together with the experimental
analysis of Ref.~\cite{LHCb:2021auc} is compatible with
that of Ref.~\cite{Janc:2004qn}, which underlined the importance of
including both mesonic and quark degrees of freedom for the binding
and description of a tetraquark below the $D^*D$ threshold.

\section*{Acknowledgments}

We would like to thank Eulogio Oset and Feng-Kun Guo for valuable comments
on this manuscript and Mikhail Mikasenko for discussions.
M.P.V. thanks the IJCLab of Orsay, where part of this work was done,
for its hospitality.
This work is partly supported by the National Natural Science Foundation
of China under Grants No. 11735003 and No. 11975041, the Fundamental
Research Funds for the Central Universities and the Thousand
Talents Plan for Young Professionals.

\appendix
\section{Range corrections to \\ the asymptotic normalization}
\label{app:AS}

Here we calculate the range corrections to $A_S$ for the $T_{cc}^+$.
Instead of the usual method of extracting $A_S$ from the residue of
the scattering amplitude, we will consider how the effective normalization of
the wave function changes when range corrections are included.
We will begin with a single channel system and assume that range corrections
are generated by an energy-dependent contact interaction of the type
\begin{eqnarray}
  V_C^{(R)} = D\, k^2 \, \delta^{(3)}(\vec{r}\,) \, ,
\end{eqnarray}
where $D$ is a coupling, $k$ refers to the center-of-mass momentum of
the two-body system, with $k^2 = 2 \mu E_{cm}$ and
$E_{cm}$ the center-of-mass energy.
We regularize this potential with a delta-shell regulator of the type
\begin{eqnarray}
  V_C^{(R)} = \frac{D(R_c)}{4 \pi R_c^2}\,k^2\,\delta(r-R_c) \, ,
\end{eqnarray}
where $D$ is a coupling and $R_c$ a cutoff.
Fixing $D(R_c)$ to the effective range $r_e$ in two-body scattering
gives~\cite{Valderrama:2016koj}
\begin{eqnarray}
  D(R_c) = \frac{2\pi}{\mu}\,\frac{r_e}{2}\,R_c^2 + \mathcal{O}(R_c^3) \, .
\end{eqnarray}
Energy-dependent potentials change the asymptotic normalization $A_S$
in a way that is compatible with the following modified normalization
condition~\cite{Stoks:1988zz} 
\begin{eqnarray}
  1 = \int_0^{\infty} dr\, u^2 (r) \left[ 1 - 2\mu \frac{d^2}{d k^2} V(r) \right]
  \, , \label{eq:norm-energy-dependence}
\end{eqnarray}
with $u(r)$ the reduced wave function of a two-body bound state.
For a contact-range theory we have $u(r) = A_S\,e^{- \gamma r}$ and after
a few manipulations we arrive at
\begin{eqnarray}
  \frac{A_{S(C)}^2}{A_S^2} = 1- \frac{r_e}{2}\,A_{S(C)}^2 \, ,
\end{eqnarray}
for $R_c \to 0$, where $A_{S(C)} = \sqrt{2 \gamma}$ and $A_S$ are
the asymptotic normalizations in the absence and presence of
range corrections.
For a single channel problem this is equivalent to
the well-known result~\cite{Phillips:1999hh}
\begin{eqnarray}
  A_S = \sqrt{\frac{A_{S(C)}}{1 - \gamma r_e}} = \sqrt{\frac{2 \gamma}{1 - \gamma r_e}} \, .
\end{eqnarray}
The advantage of the energy-dependent potential is that we can extend
the previous result to the two-channel isospin-breaking
case directly.
In the $\{ | L \rangle, | H \rangle \}$ basis, the isospin effects can be
included by considering that the potential (or for simplicity
the coupling $D$) is a matrix in said basis
\begin{eqnarray}
  D \to
  \begin{pmatrix}
    \phantom{+}\frac{1}{2}\,D_0 + \frac{1}{2}\,D_1 &
    -\frac{1}{2}\,D_0 + \frac{1}{2}\,D_1 \\
    -\frac{1}{2}\,D_0 + \frac{1}{2}\,D_1 &
    \phantom{+}\frac{1}{2}\,D_0 + \frac{1}{2}\,D_1 & 
  \end{pmatrix} \, , \nonumber \\
\end{eqnarray}
where $D_I$ is the coupling generating the $I=0,1$ effective range.
If we extend the modified normalization condition of
Eq.~(\ref{eq:norm-energy-dependence}) to the two-channel case
with the $u_L(r) = A_S \cos{\theta_I} e^{-\gamma_L r}$ and
$u_H(r) = A_S \sin{\theta_I} e^{-\gamma_H r}$ wave functions,
we arrive at
\begin{eqnarray}
  && \frac{A_{S(C)}^2}{A_S^2} = \nonumber \\ && 1 -
  \left[ \frac{r_{e0}}{2}\,\frac{(1 - \sin{2\theta_I})}{2} +
    \frac{r_{e1}}{2}\,\frac{(1 + \sin{2\theta_I})}{2} \right]\,A_{S(C)}^2 \, ,
  \nonumber \\
\end{eqnarray}
where $r_{eI}$ refers to the effective range in the isospin channel $I=0,1$
and $A_{S(C)}$ is given by Eq.~(\ref{eq:AS-LO}).
For a molecular $T_{cc}^+$ that happens to be a pure $I=0,1$ state at
short distances (i.e. $\theta_I = \mp 45\,{}^{\circ}$),
the range corrections will simplify to
\begin{eqnarray}
  \frac{A_{S(C)}^2}{A_S^2} &=& 1 - \frac{r_{eI}}{2}\,\,A_{S(C)}^2 \, ,
\end{eqnarray}
i.e. identical to the single channel case.
Finally, it is interesting to connect the previous result with the following
often-used definition of compositeness
(check, e.g. Ref.~\cite{Sekihara:2014kya}),
\begin{eqnarray}
  X_{\rm comp} = - \sum_A | g_A |^2 \frac{d G_{0A}(E)}{d E} \Big|_{E = E_B} \, ,
\end{eqnarray}
where $A$ refers to the different two-body channels, $g_A^2$ refers to
the residue of the T-matrix at the bound state pole in the diagonal channels
($g_A^2 = \lim_{E \to E_B} (E-E_B)\,T_{AA}(E)$, with $E$ the center-of-mass energy,
$E_B$ the binding energy and $T_{AA}$ the T-matrix in the diagonal channel $AA$,
where the T-matrix is defined via
$T_{AB} = V_{AB} + \sum_C V_{AC}\,G_{0C}(E)\,T_{CB}$) and $G_{0A} = 1/(E-H_{0A})$ is
the resolvent operator for channel $A$ (with $H_{0A}$ the free Hamiltonian
for that channel).
From the previous definition, we obtain
\begin{eqnarray}
  X_{\rm comp} = \frac{A_S^2}{A_{S(C)}^2} \, ,
\end{eqnarray}
for the $T_{cc}^+$, which for a negative effective range gives $X_{\rm comp} < 1$.
For a positive effective range, the previous result will probably have to be
modified in the line of what is proposed in Ref.~\cite{Matuschek:2020gqe}
for single channel scattering.
Be it as it may, our calculations already suggest a negative effective range.


\begin{thebibliography}{63}%
\makeatletter
\providecommand \@ifxundefined [1]{%
 \@ifx{#1\undefined}
}%
\providecommand \@ifnum [1]{%
 \ifnum #1\expandafter \@firstoftwo
 \else \expandafter \@secondoftwo
 \fi
}%
\providecommand \@ifx [1]{%
 \ifx #1\expandafter \@firstoftwo
 \else \expandafter \@secondoftwo
 \fi
}%
\providecommand \natexlab [1]{#1}%
\providecommand \enquote  [1]{``#1''}%
\providecommand \bibnamefont  [1]{#1}%
\providecommand \bibfnamefont [1]{#1}%
\providecommand \citenamefont [1]{#1}%
\providecommand \href@noop [0]{\@secondoftwo}%
\providecommand \href [0]{\begingroup \@sanitize@url \@href}%
\providecommand \@href[1]{\@@startlink{#1}\@@href}%
\providecommand \@@href[1]{\endgroup#1\@@endlink}%
\providecommand \@sanitize@url [0]{\catcode `\\12\catcode `\$12\catcode
  `\&12\catcode `\#12\catcode `\^12\catcode `\_12\catcode `\%12\relax}%
\providecommand \@@startlink[1]{}%
\providecommand \@@endlink[0]{}%
\providecommand \url  [0]{\begingroup\@sanitize@url \@url }%
\providecommand \@url [1]{\endgroup\@href {#1}{\urlprefix }}%
\providecommand \urlprefix  [0]{URL }%
\providecommand \Eprint [0]{\href }%
\providecommand \doibase [0]{http://dx.doi.org/}%
\providecommand \selectlanguage [0]{\@gobble}%
\providecommand \bibinfo  [0]{\@secondoftwo}%
\providecommand \bibfield  [0]{\@secondoftwo}%
\providecommand \translation [1]{[#1]}%
\providecommand \BibitemOpen [0]{}%
\providecommand \bibitemStop [0]{}%
\providecommand \bibitemNoStop [0]{.\EOS\space}%
\providecommand \EOS [0]{\spacefactor3000\relax}%
\providecommand \BibitemShut  [1]{\csname bibitem#1\endcsname}%
\let\auto@bib@innerbib\@empty
\bibitem [{\citenamefont {Aaij}\ \emph
  {et~al.}(2021{\natexlab{a}})\citenamefont {Aaij} \emph
  {et~al.}}]{LHCb:2021vvq}%
  \BibitemOpen
  \bibfield  {author} {\bibinfo {author} {\bibfnamefont {R.}~\bibnamefont
  {Aaij}} \emph {et~al.} (\bibinfo {collaboration} {LHCb}),\ }\href@noop {} {\
  (\bibinfo {year} {2021}{\natexlab{a}})},\ \Eprint
  {http://arxiv.org/abs/2109.01038} {arXiv:2109.01038 [hep-ex]} \BibitemShut
  {NoStop}%
\bibitem [{\citenamefont {Aaij}\ \emph
  {et~al.}(2021{\natexlab{b}})\citenamefont {Aaij} \emph
  {et~al.}}]{LHCb:2021auc}%
  \BibitemOpen
  \bibfield  {author} {\bibinfo {author} {\bibfnamefont {R.}~\bibnamefont
  {Aaij}} \emph {et~al.} (\bibinfo {collaboration} {LHCb}),\ }\href@noop {} {\
  (\bibinfo {year} {2021}{\natexlab{b}})},\ \Eprint
  {http://arxiv.org/abs/2109.01056} {arXiv:2109.01056 [hep-ex]} \BibitemShut
  {NoStop}%
\bibitem [{\citenamefont {Zouzou}\ \emph {et~al.}(1986)\citenamefont {Zouzou},
  \citenamefont {Silvestre-Brac}, \citenamefont {Gignoux},\ and\ \citenamefont
  {Richard}}]{Zouzou:1986qh}%
  \BibitemOpen
  \bibfield  {author} {\bibinfo {author} {\bibfnamefont {S.}~\bibnamefont
  {Zouzou}}, \bibinfo {author} {\bibfnamefont {B.}~\bibnamefont
  {Silvestre-Brac}}, \bibinfo {author} {\bibfnamefont {C.}~\bibnamefont
  {Gignoux}}, \ and\ \bibinfo {author} {\bibfnamefont {J.~M.}\ \bibnamefont
  {Richard}},\ }\href {\doibase 10.1007/BF01557611} {\bibfield  {journal}
  {\bibinfo  {journal} {Z. Phys. C}\ }\textbf {\bibinfo {volume} {30}},\
  \bibinfo {pages} {457} (\bibinfo {year} {1986})}\BibitemShut {NoStop}%
\bibitem [{\citenamefont {Carlson}\ \emph {et~al.}(1988)\citenamefont
  {Carlson}, \citenamefont {Heller},\ and\ \citenamefont
  {Tjon}}]{Carlson:1987hh}%
  \BibitemOpen
  \bibfield  {author} {\bibinfo {author} {\bibfnamefont {J.}~\bibnamefont
  {Carlson}}, \bibinfo {author} {\bibfnamefont {L.}~\bibnamefont {Heller}}, \
  and\ \bibinfo {author} {\bibfnamefont {J.~A.}\ \bibnamefont {Tjon}},\ }\href
  {\doibase 10.1103/PhysRevD.37.744} {\bibfield  {journal} {\bibinfo  {journal}
  {Phys. Rev.}\ }\textbf {\bibinfo {volume} {D37}},\ \bibinfo {pages} {744}
  (\bibinfo {year} {1988})}\BibitemShut {NoStop}%
\bibitem [{\citenamefont {Silvestre-Brac}\ and\ \citenamefont
  {Semay}(1993)}]{Silvestre-Brac:1993zem}%
  \BibitemOpen
  \bibfield  {author} {\bibinfo {author} {\bibfnamefont {B.}~\bibnamefont
  {Silvestre-Brac}}\ and\ \bibinfo {author} {\bibfnamefont {C.}~\bibnamefont
  {Semay}},\ }\href {\doibase 10.1007/BF01565058} {\bibfield  {journal}
  {\bibinfo  {journal} {Z. Phys. C}\ }\textbf {\bibinfo {volume} {57}},\
  \bibinfo {pages} {273} (\bibinfo {year} {1993})}\BibitemShut {NoStop}%
\bibitem [{\citenamefont {Semay}\ and\ \citenamefont
  {Silvestre-Brac}(1994)}]{Semay:1994ht}%
  \BibitemOpen
  \bibfield  {author} {\bibinfo {author} {\bibfnamefont {C.}~\bibnamefont
  {Semay}}\ and\ \bibinfo {author} {\bibfnamefont {B.}~\bibnamefont
  {Silvestre-Brac}},\ }\href {\doibase 10.1007/BF01413104} {\bibfield
  {journal} {\bibinfo  {journal} {Z. Phys. C}\ }\textbf {\bibinfo {volume}
  {61}},\ \bibinfo {pages} {271} (\bibinfo {year} {1994})}\BibitemShut
  {NoStop}%
\bibitem [{\citenamefont {Pepin}\ \emph {et~al.}(1997)\citenamefont {Pepin},
  \citenamefont {Stancu}, \citenamefont {Genovese},\ and\ \citenamefont
  {Richard}}]{Pepin:1996id}%
  \BibitemOpen
  \bibfield  {author} {\bibinfo {author} {\bibfnamefont {S.}~\bibnamefont
  {Pepin}}, \bibinfo {author} {\bibfnamefont {F.}~\bibnamefont {Stancu}},
  \bibinfo {author} {\bibfnamefont {M.}~\bibnamefont {Genovese}}, \ and\
  \bibinfo {author} {\bibfnamefont {J.~M.}\ \bibnamefont {Richard}},\ }\href
  {\doibase 10.1016/S0370-2693(96)01597-3} {\bibfield  {journal} {\bibinfo
  {journal} {Phys. Lett. B}\ }\textbf {\bibinfo {volume} {393}},\ \bibinfo
  {pages} {119} (\bibinfo {year} {1997})},\ \Eprint
  {http://arxiv.org/abs/hep-ph/9609348} {arXiv:hep-ph/9609348} \BibitemShut
  {NoStop}%
\bibitem [{\citenamefont {Janc}\ and\ \citenamefont
  {Rosina}(2004)}]{Janc:2004qn}%
  \BibitemOpen
  \bibfield  {author} {\bibinfo {author} {\bibfnamefont {D.}~\bibnamefont
  {Janc}}\ and\ \bibinfo {author} {\bibfnamefont {M.}~\bibnamefont {Rosina}},\
  }\href {\doibase 10.1007/s00601-004-0068-9} {\bibfield  {journal} {\bibinfo
  {journal} {Few Body Syst.}\ }\textbf {\bibinfo {volume} {35}},\ \bibinfo
  {pages} {175} (\bibinfo {year} {2004})},\ \Eprint
  {http://arxiv.org/abs/hep-ph/0405208} {arXiv:hep-ph/0405208} \BibitemShut
  {NoStop}%
\bibitem [{\citenamefont {Navarra}\ \emph {et~al.}(2007)\citenamefont
  {Navarra}, \citenamefont {Nielsen},\ and\ \citenamefont
  {Lee}}]{Navarra:2007yw}%
  \BibitemOpen
  \bibfield  {author} {\bibinfo {author} {\bibfnamefont {F.~S.}\ \bibnamefont
  {Navarra}}, \bibinfo {author} {\bibfnamefont {M.}~\bibnamefont {Nielsen}}, \
  and\ \bibinfo {author} {\bibfnamefont {S.~H.}\ \bibnamefont {Lee}},\ }\href
  {\doibase 10.1016/j.physletb.2007.04.010} {\bibfield  {journal} {\bibinfo
  {journal} {Phys. Lett. B}\ }\textbf {\bibinfo {volume} {649}},\ \bibinfo
  {pages} {166} (\bibinfo {year} {2007})},\ \Eprint
  {http://arxiv.org/abs/hep-ph/0703071} {arXiv:hep-ph/0703071} \BibitemShut
  {NoStop}%
\bibitem [{\citenamefont {Yang}\ \emph {et~al.}(2009)\citenamefont {Yang},
  \citenamefont {Deng}, \citenamefont {Ping},\ and\ \citenamefont
  {Goldman}}]{Yang:2009zzp}%
  \BibitemOpen
  \bibfield  {author} {\bibinfo {author} {\bibfnamefont {Y.}~\bibnamefont
  {Yang}}, \bibinfo {author} {\bibfnamefont {C.}~\bibnamefont {Deng}}, \bibinfo
  {author} {\bibfnamefont {J.}~\bibnamefont {Ping}}, \ and\ \bibinfo {author}
  {\bibfnamefont {T.}~\bibnamefont {Goldman}},\ }\href {\doibase
  10.1103/PhysRevD.80.114023} {\bibfield  {journal} {\bibinfo  {journal} {Phys.
  Rev. D}\ }\textbf {\bibinfo {volume} {80}},\ \bibinfo {pages} {114023}
  (\bibinfo {year} {2009})}\BibitemShut {NoStop}%
\bibitem [{\citenamefont {Karliner}\ and\ \citenamefont
  {Rosner}(2017)}]{Karliner:2017qjm}%
  \BibitemOpen
  \bibfield  {author} {\bibinfo {author} {\bibfnamefont {M.}~\bibnamefont
  {Karliner}}\ and\ \bibinfo {author} {\bibfnamefont {J.~L.}\ \bibnamefont
  {Rosner}},\ }\href {\doibase 10.1103/PhysRevLett.119.202001} {\bibfield
  {journal} {\bibinfo  {journal} {Phys. Rev. Lett.}\ }\textbf {\bibinfo
  {volume} {119}},\ \bibinfo {pages} {202001} (\bibinfo {year} {2017})},\
  \Eprint {http://arxiv.org/abs/1707.07666} {arXiv:1707.07666 [hep-ph]}
  \BibitemShut {NoStop}%
\bibitem [{\citenamefont {Wang}\ and\ \citenamefont {Di}(2019)}]{Wang:2018poa}%
  \BibitemOpen
  \bibfield  {author} {\bibinfo {author} {\bibfnamefont {Z.-G.}\ \bibnamefont
  {Wang}}\ and\ \bibinfo {author} {\bibfnamefont {Z.-Y.}\ \bibnamefont {Di}},\
  }\href {\doibase 10.5506/APhysPolB.50.1335} {\bibfield  {journal} {\bibinfo
  {journal} {Acta Phys. Polon. B}\ }\textbf {\bibinfo {volume} {50}},\ \bibinfo
  {pages} {1335} (\bibinfo {year} {2019})},\ \Eprint
  {http://arxiv.org/abs/1807.08520} {arXiv:1807.08520 [hep-ph]} \BibitemShut
  {NoStop}%
\bibitem [{\citenamefont {Bondar}\ \emph {et~al.}(2011)\citenamefont {Bondar},
  \citenamefont {Garmash}, \citenamefont {Milstein}, \citenamefont {Mizuk},\
  and\ \citenamefont {Voloshin}}]{Bondar:2011ev}%
  \BibitemOpen
  \bibfield  {author} {\bibinfo {author} {\bibfnamefont {A.~E.}\ \bibnamefont
  {Bondar}}, \bibinfo {author} {\bibfnamefont {A.}~\bibnamefont {Garmash}},
  \bibinfo {author} {\bibfnamefont {A.~I.}\ \bibnamefont {Milstein}}, \bibinfo
  {author} {\bibfnamefont {R.}~\bibnamefont {Mizuk}}, \ and\ \bibinfo {author}
  {\bibfnamefont {M.~B.}\ \bibnamefont {Voloshin}},\ }\href {\doibase
  10.1103/PhysRevD.84.054010} {\bibfield  {journal} {\bibinfo  {journal} {Phys.
  Rev.}\ }\textbf {\bibinfo {volume} {D84}},\ \bibinfo {pages} {054010}
  (\bibinfo {year} {2011})},\ \Eprint {http://arxiv.org/abs/1105.4473}
  {arXiv:1105.4473 [hep-ph]} \BibitemShut {NoStop}%
\bibitem [{\citenamefont {Nieves}\ and\ \citenamefont
  {Valderrama}(2012)}]{Nieves:2012tt}%
  \BibitemOpen
  \bibfield  {author} {\bibinfo {author} {\bibfnamefont {J.}~\bibnamefont
  {Nieves}}\ and\ \bibinfo {author} {\bibfnamefont {M.~P.}\ \bibnamefont
  {Valderrama}},\ }\href {\doibase 10.1103/PhysRevD.86.056004} {\bibfield
  {journal} {\bibinfo  {journal} {Phys. Rev.}\ }\textbf {\bibinfo {volume}
  {D86}},\ \bibinfo {pages} {056004} (\bibinfo {year} {2012})},\ \Eprint
  {http://arxiv.org/abs/1204.2790} {arXiv:1204.2790 [hep-ph]} \BibitemShut
  {NoStop}%
\bibitem [{\citenamefont {Cincioglu}\ \emph {et~al.}(2016)\citenamefont
  {Cincioglu}, \citenamefont {Nieves}, \citenamefont {Ozpineci},\ and\
  \citenamefont {Yilmazer}}]{Cincioglu:2016fkm}%
  \BibitemOpen
  \bibfield  {author} {\bibinfo {author} {\bibfnamefont {E.}~\bibnamefont
  {Cincioglu}}, \bibinfo {author} {\bibfnamefont {J.}~\bibnamefont {Nieves}},
  \bibinfo {author} {\bibfnamefont {A.}~\bibnamefont {Ozpineci}}, \ and\
  \bibinfo {author} {\bibfnamefont {A.~U.}\ \bibnamefont {Yilmazer}},\ }\href
  {\doibase 10.1140/epjc/s10052-016-4413-1} {\bibfield  {journal} {\bibinfo
  {journal} {Eur. Phys. J.}\ }\textbf {\bibinfo {volume} {C76}},\ \bibinfo
  {pages} {576} (\bibinfo {year} {2016})},\ \Eprint
  {http://arxiv.org/abs/1606.03239} {arXiv:1606.03239 [hep-ph]} \BibitemShut
  {NoStop}%
\bibitem [{\citenamefont {Manohar}\ and\ \citenamefont
  {Wise}(1993)}]{Manohar:1992nd}%
  \BibitemOpen
  \bibfield  {author} {\bibinfo {author} {\bibfnamefont {A.~V.}\ \bibnamefont
  {Manohar}}\ and\ \bibinfo {author} {\bibfnamefont {M.~B.}\ \bibnamefont
  {Wise}},\ }\href {\doibase 10.1016/0550-3213(93)90614-U} {\bibfield
  {journal} {\bibinfo  {journal} {Nucl. Phys. B}\ }\textbf {\bibinfo {volume}
  {399}},\ \bibinfo {pages} {17} (\bibinfo {year} {1993})},\ \Eprint
  {http://arxiv.org/abs/hep-ph/9212236} {arXiv:hep-ph/9212236} \BibitemShut
  {NoStop}%
\bibitem [{\citenamefont {Tornqvist}(1994)}]{Tornqvist:1993ng}%
  \BibitemOpen
  \bibfield  {author} {\bibinfo {author} {\bibfnamefont {N.~A.}\ \bibnamefont
  {Tornqvist}},\ }\href {\doibase 10.1007/BF01413192} {\bibfield  {journal}
  {\bibinfo  {journal} {Z. Phys. C}\ }\textbf {\bibinfo {volume} {61}},\
  \bibinfo {pages} {525} (\bibinfo {year} {1994})},\ \Eprint
  {http://arxiv.org/abs/hep-ph/9310247} {arXiv:hep-ph/9310247} \BibitemShut
  {NoStop}%
\bibitem [{\citenamefont {Ericson}\ and\ \citenamefont
  {Karl}(1993)}]{Ericson:1993wy}%
  \BibitemOpen
  \bibfield  {author} {\bibinfo {author} {\bibfnamefont {T.~E.~O.}\
  \bibnamefont {Ericson}}\ and\ \bibinfo {author} {\bibfnamefont
  {G.}~\bibnamefont {Karl}},\ }\href {\doibase 10.1016/0370-2693(93)90957-J}
  {\bibfield  {journal} {\bibinfo  {journal} {Phys. Lett. B}\ }\textbf
  {\bibinfo {volume} {309}},\ \bibinfo {pages} {426} (\bibinfo {year}
  {1993})}\BibitemShut {NoStop}%
\bibitem [{\citenamefont {Molina}\ \emph {et~al.}(2010)\citenamefont {Molina},
  \citenamefont {Branz},\ and\ \citenamefont {Oset}}]{Molina:2010tx}%
  \BibitemOpen
  \bibfield  {author} {\bibinfo {author} {\bibfnamefont {R.}~\bibnamefont
  {Molina}}, \bibinfo {author} {\bibfnamefont {T.}~\bibnamefont {Branz}}, \
  and\ \bibinfo {author} {\bibfnamefont {E.}~\bibnamefont {Oset}},\ }\href
  {\doibase 10.1103/PhysRevD.82.014010} {\bibfield  {journal} {\bibinfo
  {journal} {Phys. Rev. D}\ }\textbf {\bibinfo {volume} {82}},\ \bibinfo
  {pages} {014010} (\bibinfo {year} {2010})},\ \Eprint
  {http://arxiv.org/abs/1005.0335} {arXiv:1005.0335 [hep-ph]} \BibitemShut
  {NoStop}%
\bibitem [{\citenamefont {Li}\ \emph {et~al.}(2013)\citenamefont {Li},
  \citenamefont {Sun}, \citenamefont {Liu},\ and\ \citenamefont
  {Zhu}}]{Li:2012ss}%
  \BibitemOpen
  \bibfield  {author} {\bibinfo {author} {\bibfnamefont {N.}~\bibnamefont
  {Li}}, \bibinfo {author} {\bibfnamefont {Z.-F.}\ \bibnamefont {Sun}},
  \bibinfo {author} {\bibfnamefont {X.}~\bibnamefont {Liu}}, \ and\ \bibinfo
  {author} {\bibfnamefont {S.-L.}\ \bibnamefont {Zhu}},\ }\href {\doibase
  10.1103/PhysRevD.88.114008} {\bibfield  {journal} {\bibinfo  {journal} {Phys.
  Rev. D}\ }\textbf {\bibinfo {volume} {88}},\ \bibinfo {pages} {114008}
  (\bibinfo {year} {2013})},\ \Eprint {http://arxiv.org/abs/1211.5007}
  {arXiv:1211.5007 [hep-ph]} \BibitemShut {NoStop}%
\bibitem [{\citenamefont {Liu}\ \emph {et~al.}(2019)\citenamefont {Liu},
  \citenamefont {Wu}, \citenamefont {Pavon~Valderrama}, \citenamefont {Xie},\
  and\ \citenamefont {Geng}}]{Liu:2019stu}%
  \BibitemOpen
  \bibfield  {author} {\bibinfo {author} {\bibfnamefont {M.-Z.}\ \bibnamefont
  {Liu}}, \bibinfo {author} {\bibfnamefont {T.-W.}\ \bibnamefont {Wu}},
  \bibinfo {author} {\bibfnamefont {M.}~\bibnamefont {Pavon~Valderrama}},
  \bibinfo {author} {\bibfnamefont {J.-J.}\ \bibnamefont {Xie}}, \ and\
  \bibinfo {author} {\bibfnamefont {L.-S.}\ \bibnamefont {Geng}},\ }\href
  {\doibase 10.1103/PhysRevD.99.094018} {\bibfield  {journal} {\bibinfo
  {journal} {Phys. Rev.}\ }\textbf {\bibinfo {volume} {D99}},\ \bibinfo {pages}
  {094018} (\bibinfo {year} {2019})},\ \Eprint
  {http://arxiv.org/abs/1902.03044} {arXiv:1902.03044 [hep-ph]} \BibitemShut
  {NoStop}%
\bibitem [{\citenamefont {Ding}\ \emph {et~al.}(2020)\citenamefont {Ding},
  \citenamefont {Jiang},\ and\ \citenamefont {He}}]{Ding:2020dio}%
  \BibitemOpen
  \bibfield  {author} {\bibinfo {author} {\bibfnamefont {Z.-M.}\ \bibnamefont
  {Ding}}, \bibinfo {author} {\bibfnamefont {H.-Y.}\ \bibnamefont {Jiang}}, \
  and\ \bibinfo {author} {\bibfnamefont {J.}~\bibnamefont {He}},\ }\href
  {\doibase 10.1140/epjc/s10052-020-08754-6} {\bibfield  {journal} {\bibinfo
  {journal} {Eur. Phys. J. C}\ }\textbf {\bibinfo {volume} {80}},\ \bibinfo
  {pages} {1179} (\bibinfo {year} {2020})},\ \Eprint
  {http://arxiv.org/abs/2011.04980} {arXiv:2011.04980 [hep-ph]} \BibitemShut
  {NoStop}%
\bibitem [{\citenamefont {Xu}\ \emph {et~al.}(2019)\citenamefont {Xu},
  \citenamefont {Wang}, \citenamefont {Liu},\ and\ \citenamefont
  {Liu}}]{Xu:2017tsr}%
  \BibitemOpen
  \bibfield  {author} {\bibinfo {author} {\bibfnamefont {H.}~\bibnamefont
  {Xu}}, \bibinfo {author} {\bibfnamefont {B.}~\bibnamefont {Wang}}, \bibinfo
  {author} {\bibfnamefont {Z.-W.}\ \bibnamefont {Liu}}, \ and\ \bibinfo
  {author} {\bibfnamefont {X.}~\bibnamefont {Liu}},\ }\href {\doibase
  10.1103/PhysRevD.99.014027} {\bibfield  {journal} {\bibinfo  {journal} {Phys.
  Rev. D}\ }\textbf {\bibinfo {volume} {99}},\ \bibinfo {pages} {014027}
  (\bibinfo {year} {2019})},\ \Eprint {http://arxiv.org/abs/1708.06918}
  {arXiv:1708.06918 [hep-ph]} \BibitemShut {NoStop}%
\bibitem [{\citenamefont {Li}\ \emph {et~al.}(2021)\citenamefont {Li},
  \citenamefont {Sun}, \citenamefont {Liu},\ and\ \citenamefont
  {ZHu}}]{Li:2021zbw}%
  \BibitemOpen
  \bibfield  {author} {\bibinfo {author} {\bibfnamefont {N.}~\bibnamefont
  {Li}}, \bibinfo {author} {\bibfnamefont {Z.-F.}\ \bibnamefont {Sun}},
  \bibinfo {author} {\bibfnamefont {X.}~\bibnamefont {Liu}}, \ and\ \bibinfo
  {author} {\bibfnamefont {S.-L.}\ \bibnamefont {ZHu}},\ }\href@noop {} {\
  (\bibinfo {year} {2021})},\ \Eprint {http://arxiv.org/abs/2107.13748}
  {arXiv:2107.13748 [hep-ph]} \BibitemShut {NoStop}%
\bibitem [{\citenamefont {Wu}\ \emph {et~al.}(2021)\citenamefont {Wu},
  \citenamefont {Pan}, \citenamefont {Liu}, \citenamefont {Luo}, \citenamefont
  {Liu},\ and\ \citenamefont {Geng}}]{Wu:2021kbu}%
  \BibitemOpen
  \bibfield  {author} {\bibinfo {author} {\bibfnamefont {T.-W.}\ \bibnamefont
  {Wu}}, \bibinfo {author} {\bibfnamefont {Y.-W.}\ \bibnamefont {Pan}},
  \bibinfo {author} {\bibfnamefont {M.-Z.}\ \bibnamefont {Liu}}, \bibinfo
  {author} {\bibfnamefont {S.-Q.}\ \bibnamefont {Luo}}, \bibinfo {author}
  {\bibfnamefont {X.}~\bibnamefont {Liu}}, \ and\ \bibinfo {author}
  {\bibfnamefont {L.-S.}\ \bibnamefont {Geng}},\ }\href@noop {} {\  (\bibinfo
  {year} {2021})},\ \Eprint {http://arxiv.org/abs/2108.00923} {arXiv:2108.00923
  [hep-ph]} \BibitemShut {NoStop}%
\bibitem [{\citenamefont {Agaev}\ \emph {et~al.}(2021)\citenamefont {Agaev},
  \citenamefont {Azizi},\ and\ \citenamefont {Sundu}}]{Agaev:2021vur}%
  \BibitemOpen
  \bibfield  {author} {\bibinfo {author} {\bibfnamefont {S.~S.}\ \bibnamefont
  {Agaev}}, \bibinfo {author} {\bibfnamefont {K.}~\bibnamefont {Azizi}}, \ and\
  \bibinfo {author} {\bibfnamefont {H.}~\bibnamefont {Sundu}},\ }\href@noop {}
  {\  (\bibinfo {year} {2021})},\ \Eprint {http://arxiv.org/abs/2108.00188}
  {arXiv:2108.00188 [hep-ph]} \BibitemShut {NoStop}%
\bibitem [{\citenamefont {Chen}\ \emph {et~al.}(2021)\citenamefont {Chen},
  \citenamefont {Huang}, \citenamefont {Liu},\ and\ \citenamefont
  {Zhu}}]{Chen:2021vhg}%
  \BibitemOpen
  \bibfield  {author} {\bibinfo {author} {\bibfnamefont {R.}~\bibnamefont
  {Chen}}, \bibinfo {author} {\bibfnamefont {Q.}~\bibnamefont {Huang}},
  \bibinfo {author} {\bibfnamefont {X.}~\bibnamefont {Liu}}, \ and\ \bibinfo
  {author} {\bibfnamefont {S.-L.}\ \bibnamefont {Zhu}},\ }\href@noop {} {\
  (\bibinfo {year} {2021})},\ \Eprint {http://arxiv.org/abs/2108.01911}
  {arXiv:2108.01911 [hep-ph]} \BibitemShut {NoStop}%
\bibitem [{\citenamefont {Dong}\ \emph
  {et~al.}(2021{\natexlab{a}})\citenamefont {Dong}, \citenamefont {Guo},\ and\
  \citenamefont {Zou}}]{Dong:2021bvy}%
  \BibitemOpen
  \bibfield  {author} {\bibinfo {author} {\bibfnamefont {X.-K.}\ \bibnamefont
  {Dong}}, \bibinfo {author} {\bibfnamefont {F.-K.}\ \bibnamefont {Guo}}, \
  and\ \bibinfo {author} {\bibfnamefont {B.-S.}\ \bibnamefont {Zou}},\
  }\href@noop {} {\  (\bibinfo {year} {2021}{\natexlab{a}})},\ \Eprint
  {http://arxiv.org/abs/2108.02673} {arXiv:2108.02673 [hep-ph]} \BibitemShut
  {NoStop}%
\bibitem [{\citenamefont {van Kolck}(1999)}]{vanKolck:1998bw}%
  \BibitemOpen
  \bibfield  {author} {\bibinfo {author} {\bibfnamefont {U.}~\bibnamefont {van
  Kolck}},\ }\href {\doibase 10.1016/S0375-9474(98)00612-5} {\bibfield
  {journal} {\bibinfo  {journal} {Nucl. Phys. A}\ }\textbf {\bibinfo {volume}
  {645}},\ \bibinfo {pages} {273} (\bibinfo {year} {1999})},\ \Eprint
  {http://arxiv.org/abs/nucl-th/9808007} {arXiv:nucl-th/9808007} \BibitemShut
  {NoStop}%
\bibitem [{\citenamefont {Chen}\ \emph {et~al.}(1999)\citenamefont {Chen},
  \citenamefont {Rupak},\ and\ \citenamefont {Savage}}]{Chen:1999tn}%
  \BibitemOpen
  \bibfield  {author} {\bibinfo {author} {\bibfnamefont {J.-W.}\ \bibnamefont
  {Chen}}, \bibinfo {author} {\bibfnamefont {G.}~\bibnamefont {Rupak}}, \ and\
  \bibinfo {author} {\bibfnamefont {M.~J.}\ \bibnamefont {Savage}},\ }\href
  {\doibase 10.1016/S0375-9474(99)00298-5} {\bibfield  {journal} {\bibinfo
  {journal} {Nucl. Phys. A}\ }\textbf {\bibinfo {volume} {653}},\ \bibinfo
  {pages} {386} (\bibinfo {year} {1999})},\ \Eprint
  {http://arxiv.org/abs/nucl-th/9902056} {arXiv:nucl-th/9902056} \BibitemShut
  {NoStop}%
\bibitem [{\citenamefont {Meng}\ \emph {et~al.}(2021)\citenamefont {Meng},
  \citenamefont {Wang}, \citenamefont {Wang},\ and\ \citenamefont
  {Zhu}}]{Meng:2021jnw}%
  \BibitemOpen
  \bibfield  {author} {\bibinfo {author} {\bibfnamefont {L.}~\bibnamefont
  {Meng}}, \bibinfo {author} {\bibfnamefont {G.-J.}\ \bibnamefont {Wang}},
  \bibinfo {author} {\bibfnamefont {B.}~\bibnamefont {Wang}}, \ and\ \bibinfo
  {author} {\bibfnamefont {S.-L.}\ \bibnamefont {Zhu}},\ }\href {\doibase
  10.1103/PhysRevD.104.L051502} {\bibfield  {journal} {\bibinfo  {journal}
  {Phys. Rev. D}\ }\textbf {\bibinfo {volume} {104}},\ \bibinfo {pages}
  {051502} (\bibinfo {year} {2021})},\ \Eprint
  {http://arxiv.org/abs/2107.14784} {arXiv:2107.14784 [hep-ph]} \BibitemShut
  {NoStop}%
\bibitem [{\citenamefont {Ling}\ \emph {et~al.}(2021)\citenamefont {Ling},
  \citenamefont {Liu}, \citenamefont {Geng}, \citenamefont {Wang},\ and\
  \citenamefont {Xie}}]{Ling:2021bir}%
  \BibitemOpen
  \bibfield  {author} {\bibinfo {author} {\bibfnamefont {X.-Z.}\ \bibnamefont
  {Ling}}, \bibinfo {author} {\bibfnamefont {M.-Z.}\ \bibnamefont {Liu}},
  \bibinfo {author} {\bibfnamefont {L.-S.}\ \bibnamefont {Geng}}, \bibinfo
  {author} {\bibfnamefont {E.}~\bibnamefont {Wang}}, \ and\ \bibinfo {author}
  {\bibfnamefont {J.-J.}\ \bibnamefont {Xie}},\ }\href@noop {} {\  (\bibinfo
  {year} {2021})},\ \Eprint {http://arxiv.org/abs/2108.00947} {arXiv:2108.00947
  [hep-ph]} \BibitemShut {NoStop}%
\bibitem [{\citenamefont {\textasciitilde{}Feijoo}\ \emph
  {et~al.}(2021)\citenamefont {\textasciitilde{}Feijoo}, \citenamefont
  {\textasciitilde{}Liang},\ and\ \citenamefont
  {\textasciitilde{}Oset}}]{Feijoo:2021ppq}%
  \BibitemOpen
  \bibfield  {author} {\bibinfo {author} {\bibfnamefont {A.}~\bibnamefont
  {\textasciitilde{}Feijoo}}, \bibinfo {author} {\bibfnamefont {W.~t.}\
  \bibnamefont {\textasciitilde{}Liang}}, \ and\ \bibinfo {author}
  {\bibfnamefont {E.}~\bibnamefont {\textasciitilde{}Oset}},\ }\href@noop {} {\
   (\bibinfo {year} {2021})},\ \Eprint {http://arxiv.org/abs/2108.02730}
  {arXiv:2108.02730 [hep-ph]} \BibitemShut {NoStop}%
\bibitem [{\citenamefont {Qin}\ and\ \citenamefont {Yu}(2020)}]{Qin:2020zlg}%
  \BibitemOpen
  \bibfield  {author} {\bibinfo {author} {\bibfnamefont {Q.}~\bibnamefont
  {Qin}}\ and\ \bibinfo {author} {\bibfnamefont {F.-S.}\ \bibnamefont {Yu}},\
  }\href {\doibase 10.1088/1674-1137/ac1b97} {\  (\bibinfo {year} {2020}),\
  10.1088/1674-1137/ac1b97},\ \Eprint {http://arxiv.org/abs/2008.08026}
  {arXiv:2008.08026 [hep-ph]} \BibitemShut {NoStop}%
\bibitem [{\citenamefont {Feng}\ \emph {et~al.}(2013)\citenamefont {Feng},
  \citenamefont {Guo},\ and\ \citenamefont {Zou}}]{Feng:2013kea}%
  \BibitemOpen
  \bibfield  {author} {\bibinfo {author} {\bibfnamefont {G.~Q.}\ \bibnamefont
  {Feng}}, \bibinfo {author} {\bibfnamefont {X.~H.}\ \bibnamefont {Guo}}, \
  and\ \bibinfo {author} {\bibfnamefont {B.~S.}\ \bibnamefont {Zou}},\
  }\href@noop {} {\  (\bibinfo {year} {2013})},\ \Eprint
  {http://arxiv.org/abs/1309.7813} {arXiv:1309.7813 [hep-ph]} \BibitemShut
  {NoStop}%
\bibitem [{\citenamefont {Deng}\ \emph {et~al.}(2020)\citenamefont {Deng},
  \citenamefont {Chen},\ and\ \citenamefont {Ping}}]{Deng:2018kly}%
  \BibitemOpen
  \bibfield  {author} {\bibinfo {author} {\bibfnamefont {C.}~\bibnamefont
  {Deng}}, \bibinfo {author} {\bibfnamefont {H.}~\bibnamefont {Chen}}, \ and\
  \bibinfo {author} {\bibfnamefont {J.}~\bibnamefont {Ping}},\ }\href {\doibase
  10.1140/epja/s10050-019-00012-y} {\bibfield  {journal} {\bibinfo  {journal}
  {Eur. Phys. J. A}\ }\textbf {\bibinfo {volume} {56}},\ \bibinfo {pages} {9}
  (\bibinfo {year} {2020})},\ \Eprint {http://arxiv.org/abs/1811.06462}
  {arXiv:1811.06462 [hep-ph]} \BibitemShut {NoStop}%
\bibitem [{\citenamefont {Yang}\ \emph {et~al.}(2020)\citenamefont {Yang},
  \citenamefont {Ping},\ and\ \citenamefont {Segovia}}]{Yang:2019itm}%
  \BibitemOpen
  \bibfield  {author} {\bibinfo {author} {\bibfnamefont {G.}~\bibnamefont
  {Yang}}, \bibinfo {author} {\bibfnamefont {J.}~\bibnamefont {Ping}}, \ and\
  \bibinfo {author} {\bibfnamefont {J.}~\bibnamefont {Segovia}},\ }\href
  {\doibase 10.1103/PhysRevD.101.014001} {\bibfield  {journal} {\bibinfo
  {journal} {Phys. Rev. D}\ }\textbf {\bibinfo {volume} {101}},\ \bibinfo
  {pages} {014001} (\bibinfo {year} {2020})},\ \Eprint
  {http://arxiv.org/abs/1911.00215} {arXiv:1911.00215 [hep-ph]} \BibitemShut
  {NoStop}%
\bibitem [{\citenamefont {Gao}\ \emph {et~al.}(2020)\citenamefont {Gao},
  \citenamefont {Jia}, \citenamefont {Sun}, \citenamefont {Zhang},
  \citenamefont {Liu},\ and\ \citenamefont {Mei}}]{Gao:2020ogo}%
  \BibitemOpen
  \bibfield  {author} {\bibinfo {author} {\bibfnamefont {D.}~\bibnamefont
  {Gao}}, \bibinfo {author} {\bibfnamefont {D.}~\bibnamefont {Jia}}, \bibinfo
  {author} {\bibfnamefont {Y.-J.}\ \bibnamefont {Sun}}, \bibinfo {author}
  {\bibfnamefont {Z.}~\bibnamefont {Zhang}}, \bibinfo {author} {\bibfnamefont
  {W.-N.}\ \bibnamefont {Liu}}, \ and\ \bibinfo {author} {\bibfnamefont
  {Q.}~\bibnamefont {Mei}},\ }\href@noop {} {\  (\bibinfo {year} {2020})},\
  \Eprint {http://arxiv.org/abs/2007.15213} {arXiv:2007.15213 [hep-ph]}
  \BibitemShut {NoStop}%
\bibitem [{\citenamefont {Gelman}\ and\ \citenamefont
  {Nussinov}(2003)}]{Gelman:2002wf}%
  \BibitemOpen
  \bibfield  {author} {\bibinfo {author} {\bibfnamefont {B.~A.}\ \bibnamefont
  {Gelman}}\ and\ \bibinfo {author} {\bibfnamefont {S.}~\bibnamefont
  {Nussinov}},\ }\href {\doibase 10.1016/S0370-2693(02)03069-1} {\bibfield
  {journal} {\bibinfo  {journal} {Phys. Lett. B}\ }\textbf {\bibinfo {volume}
  {551}},\ \bibinfo {pages} {296} (\bibinfo {year} {2003})},\ \Eprint
  {http://arxiv.org/abs/hep-ph/0209095} {arXiv:hep-ph/0209095} \BibitemShut
  {NoStop}%
\bibitem [{\citenamefont {Vijande}\ \emph {et~al.}(2004)\citenamefont
  {Vijande}, \citenamefont {Fernandez}, \citenamefont {Valcarce},\ and\
  \citenamefont {Silvestre-Brac}}]{Vijande:2003ki}%
  \BibitemOpen
  \bibfield  {author} {\bibinfo {author} {\bibfnamefont {J.}~\bibnamefont
  {Vijande}}, \bibinfo {author} {\bibfnamefont {F.}~\bibnamefont {Fernandez}},
  \bibinfo {author} {\bibfnamefont {A.}~\bibnamefont {Valcarce}}, \ and\
  \bibinfo {author} {\bibfnamefont {B.}~\bibnamefont {Silvestre-Brac}},\ }\href
  {\doibase 10.1140/epja/i2003-10128-9} {\bibfield  {journal} {\bibinfo
  {journal} {Eur. Phys. J. A}\ }\textbf {\bibinfo {volume} {19}},\ \bibinfo
  {pages} {383} (\bibinfo {year} {2004})},\ \Eprint
  {http://arxiv.org/abs/hep-ph/0310007} {arXiv:hep-ph/0310007} \BibitemShut
  {NoStop}%
\bibitem [{\citenamefont {Lee}\ and\ \citenamefont {Yasui}(2009)}]{Lee:2009rt}%
  \BibitemOpen
  \bibfield  {author} {\bibinfo {author} {\bibfnamefont {S.~H.}\ \bibnamefont
  {Lee}}\ and\ \bibinfo {author} {\bibfnamefont {S.}~\bibnamefont {Yasui}},\
  }\href {\doibase 10.1140/epjc/s10052-009-1140-x} {\bibfield  {journal}
  {\bibinfo  {journal} {Eur. Phys. J. C}\ }\textbf {\bibinfo {volume} {64}},\
  \bibinfo {pages} {283} (\bibinfo {year} {2009})},\ \Eprint
  {http://arxiv.org/abs/0901.2977} {arXiv:0901.2977 [hep-ph]} \BibitemShut
  {NoStop}%
\bibitem [{\citenamefont {Ikeda}\ \emph {et~al.}(2014)\citenamefont {Ikeda},
  \citenamefont {Charron}, \citenamefont {Aoki}, \citenamefont {Doi},
  \citenamefont {Hatsuda}, \citenamefont {Inoue}, \citenamefont {Ishii},
  \citenamefont {Murano}, \citenamefont {Nemura},\ and\ \citenamefont
  {Sasaki}}]{Ikeda:2013vwa}%
  \BibitemOpen
  \bibfield  {author} {\bibinfo {author} {\bibfnamefont {Y.}~\bibnamefont
  {Ikeda}}, \bibinfo {author} {\bibfnamefont {B.}~\bibnamefont {Charron}},
  \bibinfo {author} {\bibfnamefont {S.}~\bibnamefont {Aoki}}, \bibinfo {author}
  {\bibfnamefont {T.}~\bibnamefont {Doi}}, \bibinfo {author} {\bibfnamefont
  {T.}~\bibnamefont {Hatsuda}}, \bibinfo {author} {\bibfnamefont
  {T.}~\bibnamefont {Inoue}}, \bibinfo {author} {\bibfnamefont
  {N.}~\bibnamefont {Ishii}}, \bibinfo {author} {\bibfnamefont
  {K.}~\bibnamefont {Murano}}, \bibinfo {author} {\bibfnamefont
  {H.}~\bibnamefont {Nemura}}, \ and\ \bibinfo {author} {\bibfnamefont
  {K.}~\bibnamefont {Sasaki}},\ }\href {\doibase
  10.1016/j.physletb.2014.01.002} {\bibfield  {journal} {\bibinfo  {journal}
  {Phys. Lett. B}\ }\textbf {\bibinfo {volume} {729}},\ \bibinfo {pages} {85}
  (\bibinfo {year} {2014})},\ \Eprint {http://arxiv.org/abs/1311.6214}
  {arXiv:1311.6214 [hep-lat]} \BibitemShut {NoStop}%
\bibitem [{\citenamefont {Junnarkar}\ \emph {et~al.}(2019)\citenamefont
  {Junnarkar}, \citenamefont {Mathur},\ and\ \citenamefont
  {Padmanath}}]{Junnarkar:2018twb}%
  \BibitemOpen
  \bibfield  {author} {\bibinfo {author} {\bibfnamefont {P.}~\bibnamefont
  {Junnarkar}}, \bibinfo {author} {\bibfnamefont {N.}~\bibnamefont {Mathur}}, \
  and\ \bibinfo {author} {\bibfnamefont {M.}~\bibnamefont {Padmanath}},\ }\href
  {\doibase 10.1103/PhysRevD.99.034507} {\bibfield  {journal} {\bibinfo
  {journal} {Phys. Rev. D}\ }\textbf {\bibinfo {volume} {99}},\ \bibinfo
  {pages} {034507} (\bibinfo {year} {2019})},\ \Eprint
  {http://arxiv.org/abs/1810.12285} {arXiv:1810.12285 [hep-lat]} \BibitemShut
  {NoStop}%
\bibitem [{\citenamefont {Fleming}\ \emph {et~al.}(2021)\citenamefont
  {Fleming}, \citenamefont {Hodges},\ and\ \citenamefont
  {Mehen}}]{Fleming:2021wmk}%
  \BibitemOpen
  \bibfield  {author} {\bibinfo {author} {\bibfnamefont {S.}~\bibnamefont
  {Fleming}}, \bibinfo {author} {\bibfnamefont {R.}~\bibnamefont {Hodges}}, \
  and\ \bibinfo {author} {\bibfnamefont {T.}~\bibnamefont {Mehen}},\
  }\href@noop {} {\  (\bibinfo {year} {2021})},\ \Eprint
  {http://arxiv.org/abs/2109.02188} {arXiv:2109.02188 [hep-ph]} \BibitemShut
  {NoStop}%
\bibitem [{\citenamefont {Hanhart}\ \emph {et~al.}(2015)\citenamefont
  {Hanhart}, \citenamefont {Kalashnikova}, \citenamefont {Matuschek},
  \citenamefont {Mizuk}, \citenamefont {Nefediev},\ and\ \citenamefont
  {Wang}}]{Hanhart:2015cua}%
  \BibitemOpen
  \bibfield  {author} {\bibinfo {author} {\bibfnamefont {C.}~\bibnamefont
  {Hanhart}}, \bibinfo {author} {\bibfnamefont {Y.~S.}\ \bibnamefont
  {Kalashnikova}}, \bibinfo {author} {\bibfnamefont {P.}~\bibnamefont
  {Matuschek}}, \bibinfo {author} {\bibfnamefont {R.~V.}\ \bibnamefont
  {Mizuk}}, \bibinfo {author} {\bibfnamefont {A.~V.}\ \bibnamefont {Nefediev}},
  \ and\ \bibinfo {author} {\bibfnamefont {Q.}~\bibnamefont {Wang}},\ }\href
  {\doibase 10.1103/PhysRevLett.115.202001} {\bibfield  {journal} {\bibinfo
  {journal} {Phys. Rev. Lett.}\ }\textbf {\bibinfo {volume} {115}},\ \bibinfo
  {pages} {202001} (\bibinfo {year} {2015})},\ \Eprint
  {http://arxiv.org/abs/1507.00382} {arXiv:1507.00382 [hep-ph]} \BibitemShut
  {NoStop}%
\bibitem [{\citenamefont {Dong}\ \emph
  {et~al.}(2021{\natexlab{b}})\citenamefont {Dong}, \citenamefont {Guo},\ and\
  \citenamefont {Zou}}]{Dong:2020hxe}%
  \BibitemOpen
  \bibfield  {author} {\bibinfo {author} {\bibfnamefont {X.-K.}\ \bibnamefont
  {Dong}}, \bibinfo {author} {\bibfnamefont {F.-K.}\ \bibnamefont {Guo}}, \
  and\ \bibinfo {author} {\bibfnamefont {B.-S.}\ \bibnamefont {Zou}},\ }\href
  {\doibase 10.1103/PhysRevLett.126.152001} {\bibfield  {journal} {\bibinfo
  {journal} {Phys. Rev. Lett.}\ }\textbf {\bibinfo {volume} {126}},\ \bibinfo
  {pages} {152001} (\bibinfo {year} {2021}{\natexlab{b}})},\ \Eprint
  {http://arxiv.org/abs/2011.14517} {arXiv:2011.14517 [hep-ph]} \BibitemShut
  {NoStop}%
\bibitem [{\citenamefont {Weinberg}(1990)}]{Weinberg:1990rz}%
  \BibitemOpen
  \bibfield  {author} {\bibinfo {author} {\bibfnamefont {S.}~\bibnamefont
  {Weinberg}},\ }\href@noop {} {\bibfield  {journal} {\bibinfo  {journal}
  {Phys. Lett.}\ }\textbf {\bibinfo {volume} {B251}},\ \bibinfo {pages} {288}
  (\bibinfo {year} {1990})}\BibitemShut {NoStop}%
\bibitem [{\citenamefont {Weinberg}(1991)}]{Weinberg:1991um}%
  \BibitemOpen
  \bibfield  {author} {\bibinfo {author} {\bibfnamefont {S.}~\bibnamefont
  {Weinberg}},\ }\href@noop {} {\bibfield  {journal} {\bibinfo  {journal}
  {Nucl. Phys.}\ }\textbf {\bibinfo {volume} {B363}},\ \bibinfo {pages} {3}
  (\bibinfo {year} {1991})}\BibitemShut {NoStop}%
\bibitem [{\citenamefont {Valderrama}(2012)}]{Valderrama:2012jv}%
  \BibitemOpen
  \bibfield  {author} {\bibinfo {author} {\bibfnamefont {M.~P.}\ \bibnamefont
  {Valderrama}},\ }\href {\doibase 10.1103/PhysRevD.85.114037} {\bibfield
  {journal} {\bibinfo  {journal} {Phys. Rev.}\ }\textbf {\bibinfo {volume}
  {D85}},\ \bibinfo {pages} {114037} (\bibinfo {year} {2012})},\ \Eprint
  {http://arxiv.org/abs/1204.2400} {arXiv:1204.2400 [hep-ph]} \BibitemShut
  {NoStop}%
\bibitem [{\citenamefont {Pav\'on~Valderrama}\ and\ \citenamefont
  {Phillips}(2015)}]{PavonValderrama:2014zeq}%
  \BibitemOpen
  \bibfield  {author} {\bibinfo {author} {\bibfnamefont {M.}~\bibnamefont
  {Pav\'on~Valderrama}}\ and\ \bibinfo {author} {\bibfnamefont {D.~R.}\
  \bibnamefont {Phillips}},\ }\href {\doibase 10.1103/PhysRevLett.114.082502}
  {\bibfield  {journal} {\bibinfo  {journal} {Phys. Rev. Lett.}\ }\textbf
  {\bibinfo {volume} {114}},\ \bibinfo {pages} {082502} (\bibinfo {year}
  {2015})},\ \Eprint {http://arxiv.org/abs/1407.0437} {arXiv:1407.0437
  [nucl-th]} \BibitemShut {NoStop}%
\bibitem [{\citenamefont {Fleming}\ \emph {et~al.}(2007)\citenamefont
  {Fleming}, \citenamefont {Kusunoki}, \citenamefont {Mehen},\ and\
  \citenamefont {van Kolck}}]{Fleming:2007rp}%
  \BibitemOpen
  \bibfield  {author} {\bibinfo {author} {\bibfnamefont {S.}~\bibnamefont
  {Fleming}}, \bibinfo {author} {\bibfnamefont {M.}~\bibnamefont {Kusunoki}},
  \bibinfo {author} {\bibfnamefont {T.}~\bibnamefont {Mehen}}, \ and\ \bibinfo
  {author} {\bibfnamefont {U.}~\bibnamefont {van Kolck}},\ }\href {\doibase
  10.1103/PhysRevD.76.034006} {\bibfield  {journal} {\bibinfo  {journal} {Phys.
  Rev.}\ }\textbf {\bibinfo {volume} {D76}},\ \bibinfo {pages} {034006}
  (\bibinfo {year} {2007})},\ \Eprint {http://arxiv.org/abs/hep-ph/0703168}
  {arXiv:hep-ph/0703168} \BibitemShut {NoStop}%
\bibitem [{\citenamefont {Dai}\ \emph {et~al.}(2020)\citenamefont {Dai},
  \citenamefont {Guo},\ and\ \citenamefont {Mehen}}]{Dai:2019hrf}%
  \BibitemOpen
  \bibfield  {author} {\bibinfo {author} {\bibfnamefont {L.}~\bibnamefont
  {Dai}}, \bibinfo {author} {\bibfnamefont {F.-K.}\ \bibnamefont {Guo}}, \ and\
  \bibinfo {author} {\bibfnamefont {T.}~\bibnamefont {Mehen}},\ }\href
  {\doibase 10.1103/PhysRevD.101.054024} {\bibfield  {journal} {\bibinfo
  {journal} {Phys. Rev. D}\ }\textbf {\bibinfo {volume} {101}},\ \bibinfo
  {pages} {054024} (\bibinfo {year} {2020})},\ \Eprint
  {http://arxiv.org/abs/1912.04317} {arXiv:1912.04317 [hep-ph]} \BibitemShut
  {NoStop}%
\bibitem [{\citenamefont {Beane}\ and\ \citenamefont
  {Savage}(2001)}]{Beane:2000fi}%
  \BibitemOpen
  \bibfield  {author} {\bibinfo {author} {\bibfnamefont {S.~R.}\ \bibnamefont
  {Beane}}\ and\ \bibinfo {author} {\bibfnamefont {M.~J.}\ \bibnamefont
  {Savage}},\ }\href {\doibase 10.1016/S0375-9474(01)01088-0} {\bibfield
  {journal} {\bibinfo  {journal} {Nucl. Phys. A}\ }\textbf {\bibinfo {volume}
  {694}},\ \bibinfo {pages} {511} (\bibinfo {year} {2001})},\ \Eprint
  {http://arxiv.org/abs/nucl-th/0011067} {arXiv:nucl-th/0011067} \BibitemShut
  {NoStop}%
\bibitem [{\citenamefont {Phillips}\ \emph {et~al.}(2000)\citenamefont
  {Phillips}, \citenamefont {Rupak},\ and\ \citenamefont
  {Savage}}]{Phillips:1999hh}%
  \BibitemOpen
  \bibfield  {author} {\bibinfo {author} {\bibfnamefont {D.~R.}\ \bibnamefont
  {Phillips}}, \bibinfo {author} {\bibfnamefont {G.}~\bibnamefont {Rupak}}, \
  and\ \bibinfo {author} {\bibfnamefont {M.~J.}\ \bibnamefont {Savage}},\
  }\href {\doibase 10.1016/S0370-2693(99)01496-3} {\bibfield  {journal}
  {\bibinfo  {journal} {Phys. Lett. B}\ }\textbf {\bibinfo {volume} {473}},\
  \bibinfo {pages} {209} (\bibinfo {year} {2000})},\ \Eprint
  {http://arxiv.org/abs/nucl-th/9908054} {arXiv:nucl-th/9908054} \BibitemShut
  {NoStop}%
\bibitem [{\citenamefont {Zyla}\ \emph {et~al.}(2020)\citenamefont {Zyla} \emph
  {et~al.}}]{Zyla:2020zbs}%
  \BibitemOpen
  \bibfield  {author} {\bibinfo {author} {\bibfnamefont {P.}~\bibnamefont
  {Zyla}} \emph {et~al.} (\bibinfo {collaboration} {Particle Data Group}),\
  }\href {\doibase 10.1093/ptep/ptaa104} {\bibfield  {journal} {\bibinfo
  {journal} {PTEP}\ }\textbf {\bibinfo {volume} {2020}},\ \bibinfo {pages}
  {083C01} (\bibinfo {year} {2020})}\BibitemShut {NoStop}%
\bibitem [{\citenamefont {Detmold}\ \emph {et~al.}(2007)\citenamefont
  {Detmold}, \citenamefont {Orginos},\ and\ \citenamefont
  {Savage}}]{Detmold:2007wk}%
  \BibitemOpen
  \bibfield  {author} {\bibinfo {author} {\bibfnamefont {W.}~\bibnamefont
  {Detmold}}, \bibinfo {author} {\bibfnamefont {K.}~\bibnamefont {Orginos}}, \
  and\ \bibinfo {author} {\bibfnamefont {M.~J.}\ \bibnamefont {Savage}},\
  }\href {\doibase 10.1103/PhysRevD.76.114503} {\bibfield  {journal} {\bibinfo
  {journal} {Phys. Rev.}\ }\textbf {\bibinfo {volume} {D76}},\ \bibinfo {pages}
  {114503} (\bibinfo {year} {2007})},\ \Eprint
  {http://arxiv.org/abs/hep-lat/0703009} {arXiv:hep-lat/0703009 [HEP-LAT]}
  \BibitemShut {NoStop}%
\bibitem [{\citenamefont {Wang}\ \emph {et~al.}(2019)\citenamefont {Wang},
  \citenamefont {Liu},\ and\ \citenamefont {Liu}}]{Wang:2018atz}%
  \BibitemOpen
  \bibfield  {author} {\bibinfo {author} {\bibfnamefont {B.}~\bibnamefont
  {Wang}}, \bibinfo {author} {\bibfnamefont {Z.-W.}\ \bibnamefont {Liu}}, \
  and\ \bibinfo {author} {\bibfnamefont {X.}~\bibnamefont {Liu}},\ }\href
  {\doibase 10.1103/PhysRevD.99.036007} {\bibfield  {journal} {\bibinfo
  {journal} {Phys. Rev. D}\ }\textbf {\bibinfo {volume} {99}},\ \bibinfo
  {pages} {036007} (\bibinfo {year} {2019})},\ \Eprint
  {http://arxiv.org/abs/1812.04457} {arXiv:1812.04457 [hep-ph]} \BibitemShut
  {NoStop}%
\bibitem [{\citenamefont {Guo}\ \emph {et~al.}(2014)\citenamefont {Guo},
  \citenamefont {Hidalgo-Duque}, \citenamefont {Nieves}, \citenamefont
  {Ozpineci},\ and\ \citenamefont {Valderrama}}]{Guo:2014hqa}%
  \BibitemOpen
  \bibfield  {author} {\bibinfo {author} {\bibfnamefont {F.~K.}\ \bibnamefont
  {Guo}}, \bibinfo {author} {\bibfnamefont {C.}~\bibnamefont {Hidalgo-Duque}},
  \bibinfo {author} {\bibfnamefont {J.}~\bibnamefont {Nieves}}, \bibinfo
  {author} {\bibfnamefont {A.}~\bibnamefont {Ozpineci}}, \ and\ \bibinfo
  {author} {\bibfnamefont {M.~P.}\ \bibnamefont {Valderrama}},\ }\href
  {\doibase 10.1140/epjc/s10052-014-2885-4} {\bibfield  {journal} {\bibinfo
  {journal} {Eur. Phys. J.}\ }\textbf {\bibinfo {volume} {C74}},\ \bibinfo
  {pages} {2885} (\bibinfo {year} {2014})},\ \Eprint
  {http://arxiv.org/abs/1404.1776} {arXiv:1404.1776 [hep-ph]} \BibitemShut
  {NoStop}%
\bibitem [{\citenamefont {de~Swart}\ \emph {et~al.}(1995)\citenamefont
  {de~Swart}, \citenamefont {Terheggen},\ and\ \citenamefont
  {Stoks}}]{deSwart:1995ui}%
  \BibitemOpen
  \bibfield  {author} {\bibinfo {author} {\bibfnamefont {J.~J.}\ \bibnamefont
  {de~Swart}}, \bibinfo {author} {\bibfnamefont {C.~P.~F.}\ \bibnamefont
  {Terheggen}}, \ and\ \bibinfo {author} {\bibfnamefont {V.~G.~J.}\
  \bibnamefont {Stoks}},\ }in\ \href@noop {} {\emph {\bibinfo {booktitle} {{3rd
  International Symposium on Dubna Deuteron 95}}}}\ (\bibinfo {year} {1995})\
  \Eprint {http://arxiv.org/abs/nucl-th/9509032} {arXiv:nucl-th/9509032}
  \BibitemShut {NoStop}%
\bibitem [{\citenamefont {Valderrama}(2016)}]{Valderrama:2016koj}%
  \BibitemOpen
  \bibfield  {author} {\bibinfo {author} {\bibfnamefont {M.~P.}\ \bibnamefont
  {Valderrama}},\ }\href {\doibase 10.1142/S021830131641007X} {\bibfield
  {journal} {\bibinfo  {journal} {Int. J. Mod. Phys.}\ }\textbf {\bibinfo
  {volume} {E25}},\ \bibinfo {pages} {1641007} (\bibinfo {year} {2016})},\
  \Eprint {http://arxiv.org/abs/1604.01332} {arXiv:1604.01332 [nucl-th]}
  \BibitemShut {NoStop}%
\bibitem [{\citenamefont {Stoks}\ \emph {et~al.}(1988)\citenamefont {Stoks},
  \citenamefont {van Campen}, \citenamefont {Spit},\ and\ \citenamefont
  {de~Swart}}]{Stoks:1988zz}%
  \BibitemOpen
  \bibfield  {author} {\bibinfo {author} {\bibfnamefont {V.~G.~J.}\
  \bibnamefont {Stoks}}, \bibinfo {author} {\bibfnamefont {P.~C.}\ \bibnamefont
  {van Campen}}, \bibinfo {author} {\bibfnamefont {W.}~\bibnamefont {Spit}}, \
  and\ \bibinfo {author} {\bibfnamefont {J.~J.}\ \bibnamefont {de~Swart}},\
  }\href {\doibase 10.1103/PhysRevLett.60.1932} {\bibfield  {journal} {\bibinfo
   {journal} {Phys. Rev. Lett.}\ }\textbf {\bibinfo {volume} {60}},\ \bibinfo
  {pages} {1932} (\bibinfo {year} {1988})}\BibitemShut {NoStop}%
\bibitem [{\citenamefont {Sekihara}\ \emph {et~al.}(2015)\citenamefont
  {Sekihara}, \citenamefont {Hyodo},\ and\ \citenamefont
  {Jido}}]{Sekihara:2014kya}%
  \BibitemOpen
  \bibfield  {author} {\bibinfo {author} {\bibfnamefont {T.}~\bibnamefont
  {Sekihara}}, \bibinfo {author} {\bibfnamefont {T.}~\bibnamefont {Hyodo}}, \
  and\ \bibinfo {author} {\bibfnamefont {D.}~\bibnamefont {Jido}},\ }\href
  {\doibase 10.1093/ptep/ptv081} {\bibfield  {journal} {\bibinfo  {journal}
  {PTEP}\ }\textbf {\bibinfo {volume} {2015}},\ \bibinfo {pages} {063D04}
  (\bibinfo {year} {2015})},\ \Eprint {http://arxiv.org/abs/1411.2308}
  {arXiv:1411.2308 [hep-ph]} \BibitemShut {NoStop}%
\bibitem [{\citenamefont {Matuschek}\ \emph {et~al.}(2021)\citenamefont
  {Matuschek}, \citenamefont {Baru}, \citenamefont {Guo},\ and\ \citenamefont
  {Hanhart}}]{Matuschek:2020gqe}%
  \BibitemOpen
  \bibfield  {author} {\bibinfo {author} {\bibfnamefont {I.}~\bibnamefont
  {Matuschek}}, \bibinfo {author} {\bibfnamefont {V.}~\bibnamefont {Baru}},
  \bibinfo {author} {\bibfnamefont {F.-K.}\ \bibnamefont {Guo}}, \ and\
  \bibinfo {author} {\bibfnamefont {C.}~\bibnamefont {Hanhart}},\ }\href
  {\doibase 10.1140/epja/s10050-021-00413-y} {\bibfield  {journal} {\bibinfo
  {journal} {Eur. Phys. J. A}\ }\textbf {\bibinfo {volume} {57}},\ \bibinfo
  {pages} {101} (\bibinfo {year} {2021})},\ \Eprint
  {http://arxiv.org/abs/2007.05329} {arXiv:2007.05329 [hep-ph]} \BibitemShut
  {NoStop}%
\end{thebibliography}
%

\end{document}